\documentclass[letterpaper,twocolumn,10pt]{article}
\usepackage{usenix2019_v3}

\usepackage[utf8]{inputenc}
\usepackage{xcolor}
\usepackage{xspace}
\usepackage{tikz}
\usepackage{tikzpeople}
\usepackage{cleveref}
\usepackage{enumerate}
\usepackage[normalem]{ulem}
\usepackage{booktabs}
\usepackage{multirow}
\usepackage{tabu}
\usepackage{threeparttable}
\usepackage{pifont}
\usepackage{subcaption}
\usepackage{fancyhdr}

\microtypecontext{spacing=nonfrench}

\usetikzlibrary{positioning,calc,arrows,decorations,shapes,decorations.pathreplacing,chains,fit,arrows.meta}
\tikzset{>=latex}

\pgfdeclareimage[width=2cm]{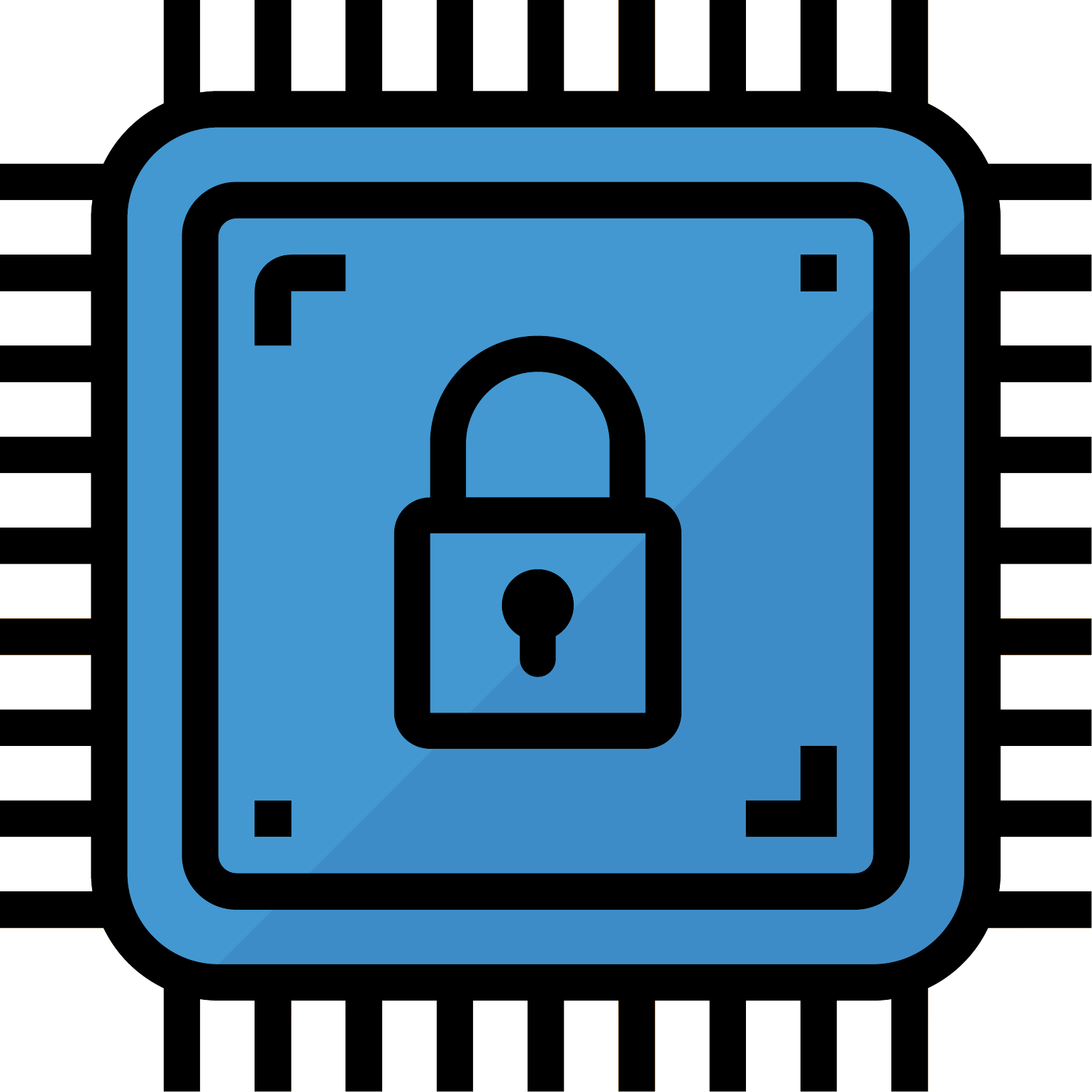}{figures/icons/enclave}
\pgfdeclareimage[width=2cm]{enclave-yellow}{figures/icons/sgx_red}
\pgfdeclareimage[width=2cm]{enclave-red}{figures/icons/sgx_yellow}
\pgfdeclareimage[width=2cm]{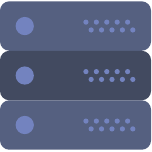}{figures/icons/server_contoured}
\pgfdeclareimage[width=2cm]{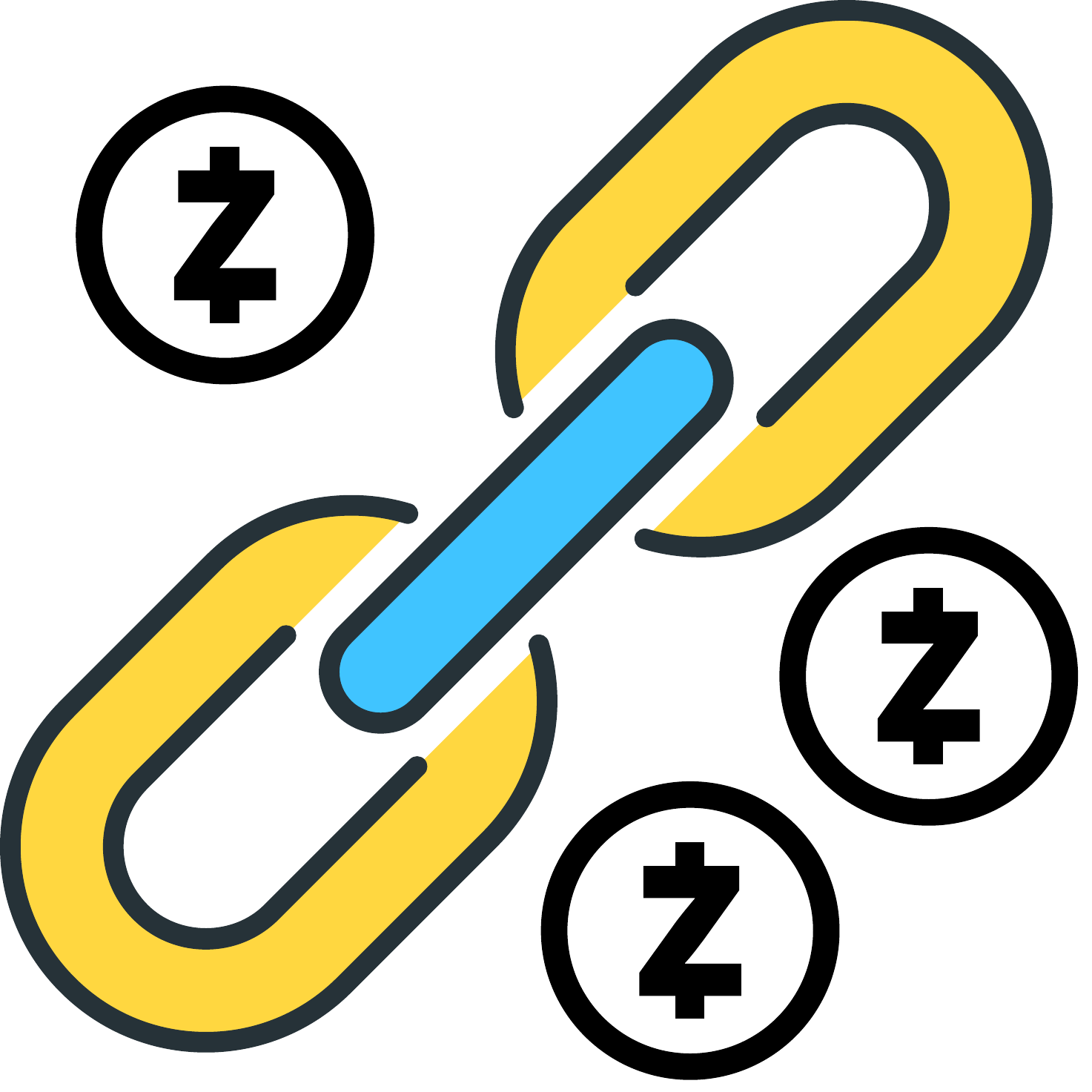}{figures/icons/blockchain}
\pgfdeclareimage[width=2cm]{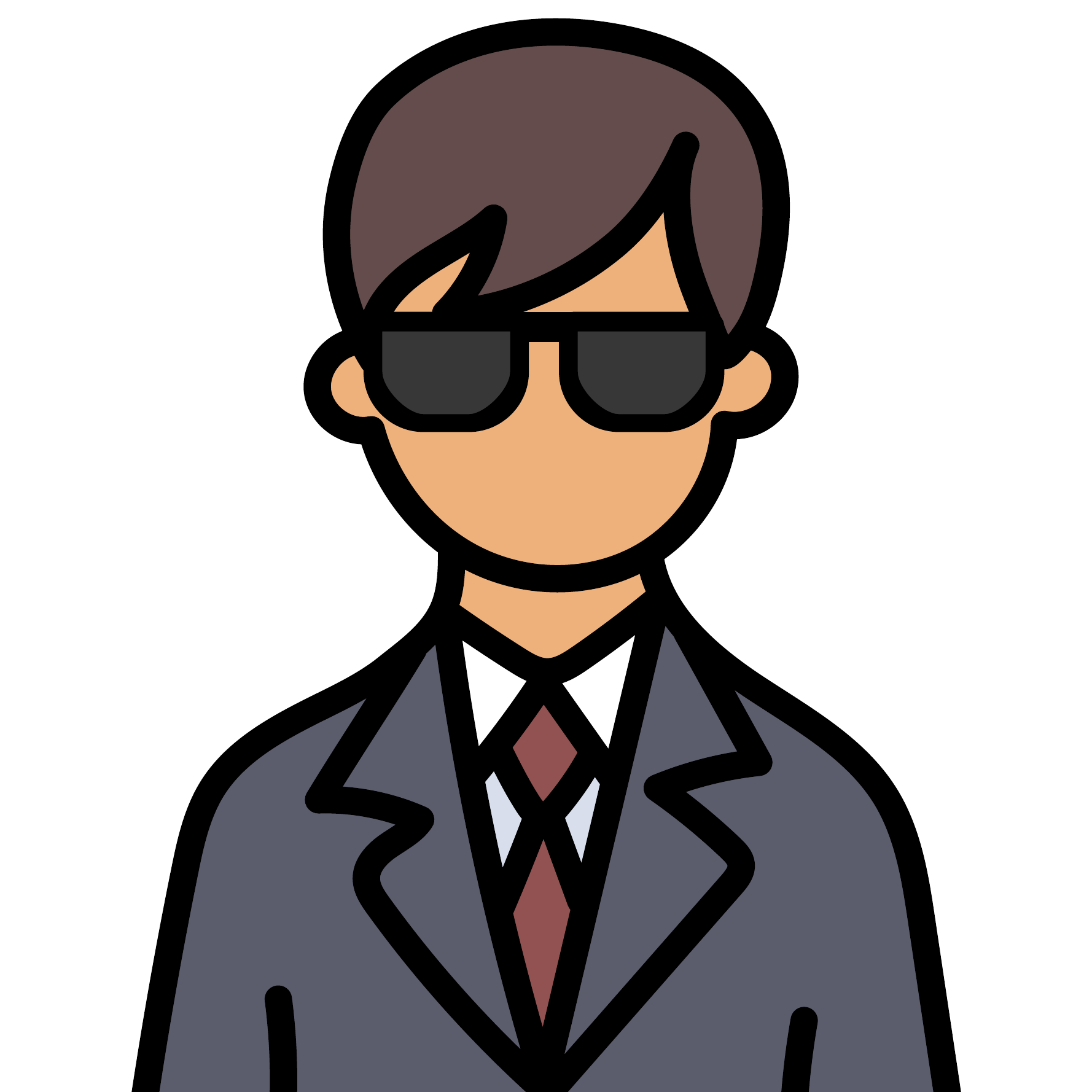}{figures/icons/renter}
\pgfdeclareimage[width=2cm]{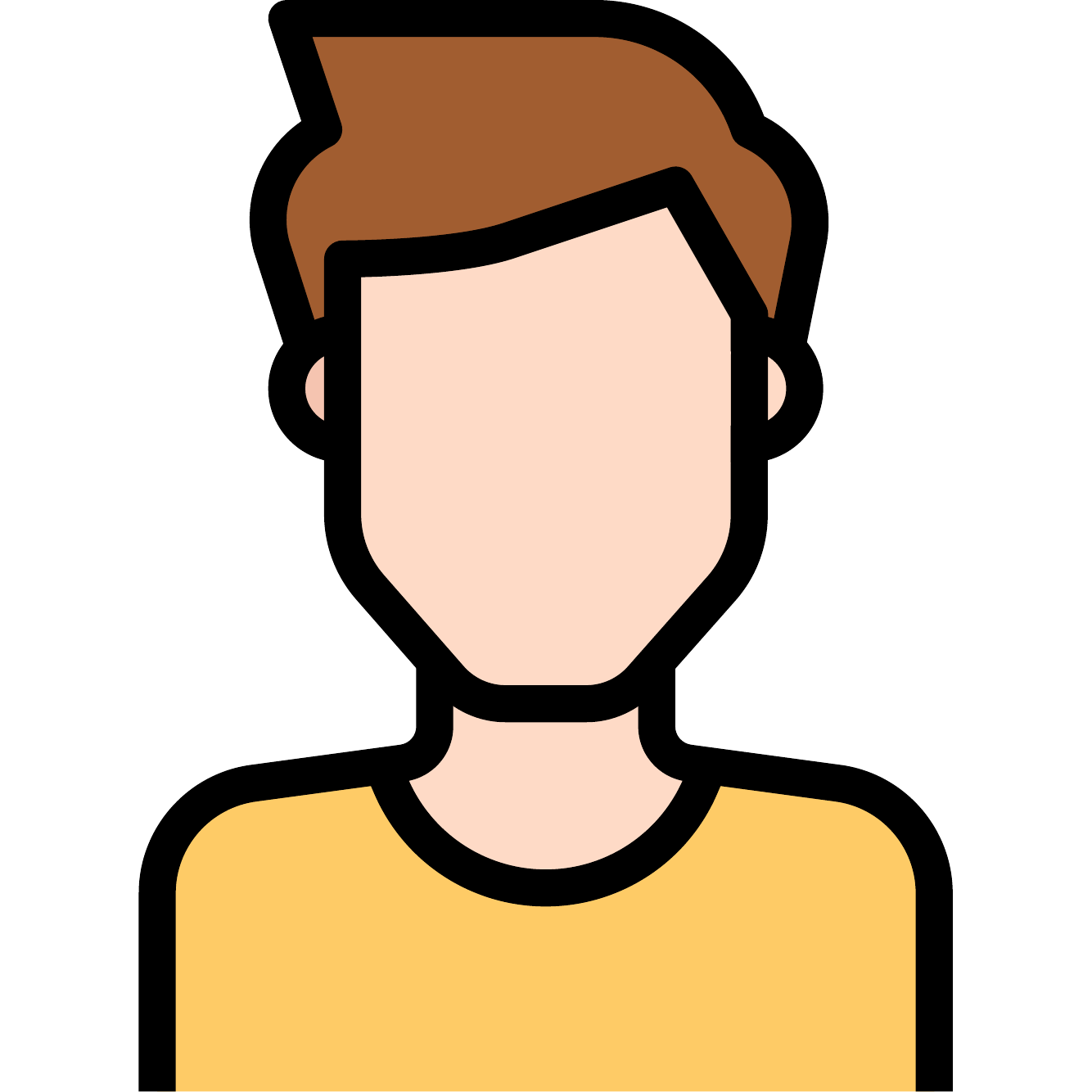}{figures/icons/owner_2}
\pgfdeclareimage[width=1cm]{phone}{figures/icons/smartphone}

\definecolor{custom-blue}{HTML}{4398d1}
\definecolor{custom-red}{HTML}{ff4842}
\definecolor{custom-yellow}{HTML}{ffe163}

\newcommand{\name}{\textsc{TEEvil}\xspace}

\newif \ifdraft
\draftfalse

\ifdraft

\newcommand{\todo}[1]{\textcolor{red}{TODO: #1}}
\newcommand{\dl}[1]{\textcolor{orange}{Daniele: #1}}
\newcommand{\ip}[1]{\textcolor{blue}{Ivan: #1}}
\newcommand{\ms}[1]{\textcolor{magenta}{Moritz: #1}}
\newcommand{\sm}[1]{\textcolor{green}{Sinisa: #1}}

\else

\newcommand{\todo}[1]{}
\newcommand{\dl}[1]{}
\newcommand{\ip}[1]{}
\newcommand{\ms}[1]{}
\newcommand{\sm}[1]{}

\fi

\renewcommand{\paragraph}{\medskip\noindent\textbf}

\newcommand{\user}{identity owner\xspace}
\newcommand{\users}{identity owners\xspace}
\newcommand{\User}{Identity owner\xspace}
\newcommand{\Users}{Identity owners\xspace}
\newcommand{\UsEr}{Identity Owner\xspace}
\newcommand{\UsErs}{Identity Owners\xspace}
\newcommand{\enclinf}{infrastructure maintainer\xspace}

\newcommand{\EnclInf}{Infrastructure Maintainer\xspace}

\newcommand{\enclinfs}{infrastructure maintainers\xspace}
\newcommand{\requester}{identity renter\xspace}
\newcommand{\requesters}{identity renters\xspace}
\newcommand{\Requester}{Identity renter\xspace}
\newcommand{\RequEster}{Identity Renter\xspace}
\newcommand{\Requesters}{Identity renters\xspace}

\ifdraft

\fi

\title{TEEvil: Identity Lease via Trusted Execution Environments}
\author{
    \begin{tabular}{c}
    {\rm Ivan \ Puddu }\\
    \end{tabular}
    &
    \begin{tabular}{c}
    {\rm Daniele \ Lain }\\
    \end{tabular}
    &
	\begin{tabular}{c}
    {\rm Moritz \ Schneider}\\
    \end{tabular}
    \\&&
    \\
    \begin{tabular}{c}
    {\rm Elizaveta \ Tretiakova }\\
    \end{tabular}
    &
    \begin{tabular}{c}
    {\rm Sinisa \ Matetic }\\
    \end{tabular}
    &
    \begin{tabular}{c}
    {\rm Srdjan \ \v{C}apkun }\\
    \end{tabular}
    \\&&
    \\
    &
    \begin{tabular}{c}
    ETH Zurich\\
    \end{tabular}
    &
}

\begin{document}

\maketitle
\ifdraft
\thispagestyle{fancy}
\fi

\begin{abstract}
We investigate \emph{identity lease}, a new type of service in which users lease their identities to third parties by providing them with full or restricted access to their online accounts or credentials. We discuss how identity lease could be abused to subvert the digital society, facilitating the spread of fake news and subverting electronic voting by enabling the sale of votes. We show that the emergence of Trusted Execution Environments and anonymous cryptocurrencies, for the first time, allows the implementation of such a lease service while guaranteeing fairness, plausible deniability and anonymity, therefore shielding the users and account renters from prosecution. To show that such a service can be practically implemented, we build an example service that we call \name leveraging Intel SGX and ZCash. Finally, we discuss defense mechanisms and challenges in the mitigation of identity lease services.

\end{abstract}

\section{Introduction}
\label{sec:introduction}

Different online platforms collect, share and monetize user's data. From this data they infer user behaviors and preferences~\cite{pennacchiotti2011machine} which can then be used for better product placement or to exert influence over the users~\cite{cambridge_analytica}. 
Not only data, but also user actions can have value and be monetized. Third parties can instruct real users to perform arbitrary actions online in exchange for money -- an activity called \textit{crowdturfing}~\cite{Wang2012surfandturf}. Crowdturfing platforms facilitate these exchanges, by connecting \emph{customers}, entities that want some action to be conducted online to \emph{workers}, regular users that can be hired to perform various tasks. These tasks range from posting advertisements, comments or reviews related to products and venues, interacting with (e.g., ``liking'') the content on Online Social Networks (OSNs), to click-fraud operations against Pay-per-Click advertisement providers. This type of advertisement is more impactful than the traditional means because it poses as ``grass-root'' word-of-mouth opinions of (allegedly) real users, whose identity lends credibility to the content~\cite{brown1987social,ferguson2008word}.

In this work we investigate a new type of user monetization, that we name \emph{identity lease}, in which users lease their accounts and online credentials to third parties. Those third parties can then control those accounts within some defined temporal and other limits. 

We show that identity lease has the potential to have a significant societal impact -- it can be used to generate and enhance the spread of fake news, and ultimately to sway elections by e.g., buying votes. 
Existing crowdturfing platforms limit the reach of such activities. Such platforms require \emph{workers} to manually execute tasks and \emph{customers} to subsequently verify them. They do not offer any privacy guarantees for the workers, and when executing payments expose the platform and workers to legal risks -- e.g., sale and purchase of votes is illegal in almost all jurisdictions~\cite{us-voter-fraud,ger-voter-fraud,ch-voter-fraud}.

To have a higher impact, such platforms either need to recruit more workers or resort to the use of fake accounts. As the example of fake news shows, the use of fake accounts can only have a limited impact~\cite{shao2017spread,twitter2016fakenews}. The fake news phenomenon is driven by targeted, carefully crafted misinformation. OSNs are then used to maximize the misinformation's impact by allowing to easily target separate audiences with different alternative facts, resulting sometimes in the change of election results~\cite{bond201261,vitak2011s,shao2017spread}. 
To make such campaigns more effective these stories should not only be started by dubious sources and spread through fake accounts but need to be accepted and spread by real users within their social networks. 
A party that could lease a large number of real identities would therefore have the ability to spread fake news more effectively and in much more subtle ways. This would amount to the purchase of a large number of small-scale influencers. 

Big social networks have been under pressure from governments to address the spread of fake news. Their natural reaction has been to track down fake news sources and fake accounts thus stopping their spread into the networks of real users~\cite{fb-fake-news,twitter2018midterms,twitter2016fakenews}. However, if fake news were to be introduced by large numbers of real active users and \emph{inserted among their regular posts} then not only would their impact be more pronounced, but it would make their identification and removal more complex for OSN operators. The identification of such carefully placed fake news would then require looking into message content~\cite{conroy2015automatic,shu2017fake} and their removal would have implications on free speech and censorship~\cite{rainie2017future}. 

Furthermore, if identity lease would be done without the manual participation of the users while preserving the confidentiality of their OSN credentials and remunerated through anonymous payments, such a system would attract a wider set of workers. 

If the platform could in addition guarantee fairness, it would then further offer all the necessary properties for the large scale sale and purchase of votes in an e-voting system. It would ensure that voters get paid only if the buyer gets valid voting credentials, as well as that the payment is executed when the vote is cast by the buyer. It would enable the voters to sell their votes while shielding them from prosecution. 

We aim to show that Trusted Execution Environments (TEEs) coupled with distributed ledgers (and thus cryptocurrencies) represent almost an ideal set of technologies that can underpin identity lease.
Ultimately, these new technologies create the ability to sell/buy/rent digital identities on a massive scale while preserving indistinguishability, plausible deniability and fairness between customers and worker, therefore, having the potential to undermine entire societies. To illustrate this, we introduce \name, a system that allows users to lease their online accounts to \emph{renters} while preserving the above properties. Further, our system can provide its users reasonable anonymity guarantees, with the assumption that some specific actions they might sell can potentially de-anonymize them.%
We implemented \name within Intel SGX~\cite{costanintel} and used ZCash~\cite{sasson2014zerocash} for anonymous payment; the renter leased a Reddit account from the user, allowing it to interact in the user's name.

This paper focuses on the technical design of an identity lease system and discusses countermeasures that could be raised against such systems, all in the hope that such behaviors would be detected and prevented before they emerge in the wild. 

\paragraph{Contributions.}

Our contributions can be summarized as follows:
\begin{itemize}
    \item \emph{Problem}: we highlight the problem that arises when a large number of people are allowed to lease their online identities and sell access to their accounts, automatically and without any repercussion. Additionally, we explore the impact of this phenomenon on existing and future e-voting systems~\cite{scytl}.
    
    \item \emph{\name System}: We present and implement \name, an architecture for identity lease that is fair, privacy preserving, and completely automated for account owners and renters. \name{} leverages a careful combination of escrows to cryptocurrencies with trusted execution to achieve fairness, and disincentivize all parties from trying to cheat the protocol. \name protects users credentials by limiting their use through TEEs~\cite{matetic2018delegatee}, and protects anonymity of such transactions through anonymous cryptocurrencies~\cite{sasson2014zerocash}. 
    In particular, \name allows the replacement of the crowdturfing middlemen with TEEs\footnote{Note that this, however, inherently shifts trust to the TEE manufacturer.}, hence removing entities from the picture that can be coerced to reveal information about owners and renters.

    \item \emph{Countermeasures}: We discuss possible countermeasures against identity lease in general, and to \name{} in particular, highlighting that stopping these systems might be challenging in some scenarios and requires more research attention. 
\end{itemize}

\paragraph{Outline.} 
The remainder of the paper is structured as follows: In Section~\ref{sec:prob_statement}, we introduce the problem, describing some applications in which an identity leasing marketplace can be disruptive in current digital societies. In Section~\ref{sec:protocol}, we introduce \name, a protocol for fair and secure identity sharing, and detail its protocol, followed by the security analysis in Section~\ref{sec:security_analysis}. Then, we change perspective and summarize possible defences that can mitigate the effects of \name in Section~\ref{sec:defences}. In Section~\ref{sec:implementation}, we describe our prototype implementation of \name and analyze its performance. In Section~\ref{sec:discussion}, we discuss two improved distributed designs of \name and how it can be combined with anonymous networks. Finally, we conclude in Section~\ref{sec:conclusions}.

\section{Problem Statement}
\label{sec:prob_statement}

In this section we introduce the problem through two motivating examples, discuss why existing techniques were not well suited to solve this problem and why TEEs and anonymous cryptocurrencies represent almost ideal candidate technologies in this problem space.

\subsection{Motivating examples}
Providing the ability to entities to lease identities on a large scale could have a number of negative consequences for digital societies. It can be used to subvert digital societies by polarizing and influencing the opinions of its members, to pollute automatic recommender systems, bootstrap the virality of content on online social networks, and subvert democratic processes, to name a few. 

We discuss two examples in more detail: electronic voting and posts in online social networks. 

\paragraph{Electronic Voting.}
Lease of government-issued electronic identities allows miscreants to buy votes in e-voting platforms. While in-person e-vote selling is unpractical (the miscreant would need to be co-located with the seller at the time of voting, as e-voting systems do not provide any proof of whom a vote was cast for), online rental would allow the miscreant to cast votes to their candidate of choice. There usually exists a subset of people which argue that their vote would not matter, and could, therefore, be persuaded to sell their vote. Even a small amount of votes can swing election results~\cite{Mulligan2003} and greatly influence politics and democracies.

\paragraph{Online Social Networks (OSNs).} 
In the context of OSNs, miscreants could leverage peer trust, and post advertisement masked as legitimate content -- for example post, ``like'', or ``reshare'' some product. Influential peers that maximize the impact of content can be selected with targeted, topic-specific heuristics~\cite{kempe2003maximizing,Barbieri2012,Chen2015,Goyal2011}. This could destroy the conventional centralized advertisement business model of OSNs, shifting to a new peer-to-peer paradigm where every account is a potential ``influencer'' since users are highly influenced by posts from their peers~\cite{ye2011influence,senecal2004influence}.
Miscreants can also manipulate content virality processes: patterns of content diffusion and virality processes on online communities~\cite{dow2013anatomy,cheng2014can,friggeri2014rumor} can be leveraged to maximize the efficacy of the crucial initial phase of the life-cycle of news and rumors~\cite{Vosoughi1146} -- a worrisome scenario for the uncontrolled spreading of fake news. 
Regarding politics, friendship relationships and trust could be leveraged to ``water down'' opposite views (for example by commenting or posting that an opposite party politician \textit{``maybe has a point''}). Such a strategy could be impactful, as diverging opinions are otherwise very unlikely to reach ``echo chambers'' that are oppositely polarized~\cite{del2016spreading}. 
Rented accounts could also give a fake sense of grass-root approval to extreme views\footnote{The virality of the ``PizzaGate'' and ``QAnon'' conspiracy theories was bootstrapped by accounts posing as real people~\cite{fisher2016pizzagate}.}, and efficiently promote fake news and conspiracy theories that find fertile soil in already polarized communities due to confirmation biases~\cite{Vosoughi1146,bessi2015viral} -- and debunking news after they have gone viral is largely ineffective~\cite{zollo2017debunking,friggeri2014rumor}.

\subsection{Requirements}
\label{sec:requirements}
A system that allows identity lease at a large scale is likely to violate the terms of service of online platforms, and in some cases will be illegal to operate or to participate in, notably in the case of a sale of votes in government elections. To achieve its goal the system therefore needs to include proper monetary incentives but at the same time guarantees fairness and impunity. Such a system would therefore need to satisfy the following properties.

\paragraph{Fairness.} 
\Users and \requester should exchange services for payment. Upon the completion of the transaction, the \requester would have used the \users's account, and the \user would have received the payment.  
There would be no incentive to put up accounts for rental on a system where fairness is not guaranteed; the same holds if \requesters are not guaranteed to receive what they pay for. 
    
\paragraph{Indistinguishability.} 
\emph{From the point of view of the service} any action coming from rented identities should be indistinguishable from actions manually performed by the original \users. 
Given the grey area of identity lease, it is likely that targeted services (for which accounts are put up for rental) would want to block such activities. If they could therefore distinguish between ``normal'' user actions and actions performed while the account is rented to a third party, they could intervene. Indistinguishability guarantees that the service is oblivious to the rental process.

\paragraph{Plausible Deniability.}
All parties should be able to plausibly deny their participation in such a system or in a particular campaign.
This would make it difficult for services or authorities to take action against \users and \requesters. 

\subsection{Existing Approaches and Limitations}
Candidate technologies that could provide the aforementioned properties include a range of fair exchange, zero-knowledge, multi-party computation, and distributed ledger techniques. However, all these approaches fail to solve this problem and to satisfy all the required properties. In fact, most would require the cooperation of online services, which goes against their business interests.

Any protocol built using the technologies mentioned above requires some proof that the commissioned actions were performed correctly. While multi-party computation (MPC) and Zero-Knowledge (ZK) protocols could be used to build such proof, they would all be limited by the interface of the service (e.g., e-voting or the OSN) to which the \user presents its credentials. These services typically offer only simple login credentials and do not run ZK or MPC protocols.%

In the example of electronic voting, a fair exchange would require the \user to either give its voting credentials to the \requester or to present the proof that he/she voted in a particular way. Deployed electronic systems do not provide such proofs (e.g.,~\cite{scytl}), and without a trusted third party, no protocol will guarantee the fairness of this exchange~\cite{even1980relations,fischer1982impossibility}.

Smart contracts~\cite{buterin2014next} that run on top of distributed ledgers can also seem like a proper solution for a fair exchange.
Juels et al. investigated how smart contracts could be used for criminal activities while still guaranteeing a fair exchange for the participants~\cite{juels-ccs16}, and Dziembowski at al. explored how general fair exchange protocols can be built on smart contracts~\cite{dziembowski2018fairswap}. However, both approaches cannot directly be applied to our setting. In~\cite{juels-ccs16}, in order to participate in the criminal smart contract, the \user needs to publicly reveal his action on the blockchain, where it will be visible by everyone, thus violating the indistinguishability and plausible deniability requirements. This limitation could be circumvented by requiring a ZK proof of the action, as done in~\cite{dziembowski2018fairswap} and for some of the use cases described in~\cite{juels-ccs16}. However, as pointed out above, this requires the cooperation of the service provider. Finally, all the current blockchains that support smart contracts and are expressive enough for our use cases do not provide any strong anonymity guarantee, making it challenging to provide plausible deniability to \users and \requesters.

A blog post by Dian et al.\ \cite{darkdao} proposed a combination of smart contracts and trusted execution environments to create a marketplace for votes in the context of blockchain voting. However, it only focuses on votes that take place entirely on the blockchain and does not easily generalize to arbitrary off-chain e-voting systems. Further, it is unclear how and whether the various proposals in the blog post can achieve fairness and plausible deniability even in the context of pure blockchain votes.

To conclude, existing techniques cannot satisfy all the properties mentioned above at the same time: so far, it is impossible to deploy a system that allows leasing identities \textit{en masse} and hence subvert digital societies.

\section{\name{}}
\label{sec:protocol}
Recently, mainstream vendors, such as Intel~\cite{costanintel}, ARM~\cite{alvestrustzone}, and RISC-V~\cite{costan2016sanctum,keystone}, started deploying Trusted Execution Environments (TEEs) in their processors. TEEs are becoming a commodity technology, increasingly widely deployed even in consumer devices. Due to their diffusion, TEEs are enabling new use-cases thanks to their security properties. However, TEEs can not only be used for good -- nefarious applications have yet to be explored. In this paper, we show how TEEs enable building a large-scale marketplace for identity lease.

There are five main parties participating in \name{}:

\begin{itemize}
    \item \textbf{\UsEr}: Any entity that enrolls in \name{} to offer some actions to be performed on their behalf.
    \item \textbf{\RequEster}: Any entity that wants to start an identity rental campaign. The \requester needs to specify what should be performed through the leased identities, and how much he would pay for such service.
    \item \textbf{Service}: The service is an online social network (OSN), e-voting system, or any online service for which a \requester is interested in using other people's identities through their accounts to perform some actions on their behalf.
    \item \textbf{\name{} Enclave:} The code running in the TEE is responsible for managing all the interactions between the \users, the \requesters, the service(s) and the funds in the cryptocurrency. A set of supported functions for the services is exposed to \requesters. Renters can then specify through an API which actions they would like to buy and for which services. The enclave performs these actions on behalf of \users and distributes the funds accordingly.
    \item \textbf{\EnclInf}: The \enclinf provides the infrastructure that hosts the enclaves. There could be multiple \enclinfs running the same services, and \requesters and \users are free to choose the ones they prefer, based for instance on the fees the \enclinfs charge or the services they support.
\end{itemize}

\name{} replaces the trusted intermediary of crowdturfing campaigns with a protocol that runs on top of TEE and anonymous cryptocurrencies on distributed ledgers. As we show, this combination of a ledger and TEEs enables large-scale identity leasing while guaranteeing fairness and plausible deniability.

The TEE provides three important properties to \name: (i) it protects the details (e.g., credentials) of the parties; (ii) it allows to automate all actions that the worker should perform (that are now performed by the verifiable code), and removes the need for manually proving and verifying the execution of actions; finally (iii) it provides fair exchange by managing funds with a decentralized anonymous cryptocurrency such as ZCash.
As it provides automatic execution of actions by the TEE from leased accounts, \name{} greatly extends crowdturfing by offering better properties, and enabling a large-scale marketplace.

In the following, we give an overview of \name, instantiating the TEE on top of which it is built with Intel SGX and using ZCash as the anonymous cryptocurrency.

\subsection{\name{} Overview}
\label{sec:overview}

\begin{figure}[tbp]
    \centering
    \resizebox{\columnwidth}{!}{
        \tikzstyle{place}=[rectangle,draw=black!50,thick,fill=white,align=center,minimum height=1cm,minimum width={{width("Advertiser")+15pt}}]
\tikzstyle{icon}=[rectangle,draw=white!50,fill=white,align=center,minimum height=1cm,minimum width=2cm]
\tikzstyle{iconsmall}=[rectangle,draw=white!50,fill=white,align=center,minimum height=1cm,minimum width=1cm]

\tikzset{every loop/.style={min distance=10mm,in=240,out=300,looseness=5}}
\begin{tikzpicture}
    \node[icon]  (E)                    {\pgfuseimage{enclave}};
    \node[icon] (U) [left=2.5cm of E]  {\pgfuseimage{owner}\\\Users};
    \node[icon] (N) [above=1.0cm of E]        {\pgfuseimage{server}\\Service};
    \node[icon] (B) [below=1.0cm of E]        {\pgfuseimage{blockchain}\\Anonymous Cryptocurrency};
    \node[icon] (V) [right=2.5cm of E] {\pgfuseimage{renter}\\\Requester};
    
    \node[iconsmall] (H) [above right=-0.2cm and -0.5cm of U] {Proxy\\\pgfuseimage{phone}};
    \node[draw=none] () [above=-0.1cm of E] {\name};

    \path[->,thick]           (U) edge node[above] {1. Credentials}     (E);
    \path[->,thick]           (B) edge node[auto] {4. Payment}         (U);
    \path[->,bend left,thick] (E) edge[out=70,in=110,looseness=2.5] node[right]  {3. Perform Action}  (N);
    \path[->,thick]           (V) edge node[above] {2a. Buy Actions}     (E);
    \path[->,thick]           (V) edge node[auto]  {2b. Provide Funds}   (B);
    \path[->,thick]           (E) edge node[fill=white]  {4. Reward user}  (B);

\end{tikzpicture}
    }
    \caption{Overview of the \name protocol showing how its entities interact with each other.}
    \label{fig:system_architecture}
\end{figure}
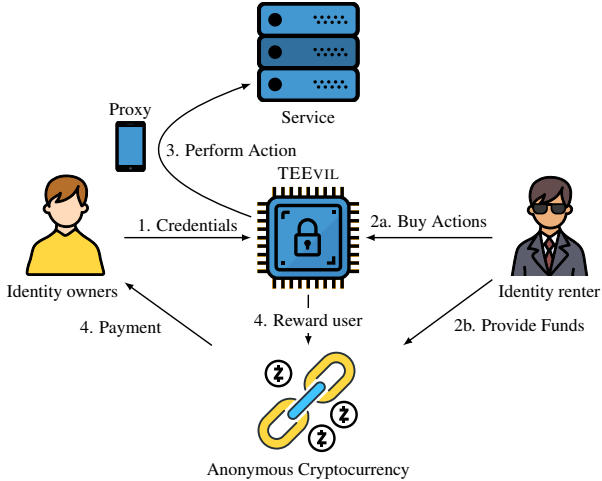

In Figure~\ref{fig:system_architecture} we show the main interactions within \name{}. \Users enroll in the system in step 1, by securely sending their credentials of the target service and their public cryptocurrency address to the \name SGX enclave. In step 2a, \requesters start a new campaign by communicating to the enclave which actions they wish to buy from \users, and in step 2b, by transferring funds to an address controlled by the enclave on an anonymous public blockchain. In step 3, the enclave automatically performs actions by using the \users' devices as proxies, so that from the service's point of view the action appears as any other normal activity of the \user. Step 3 is repeated for every user enrolled in \name. Finally, in step 4, as soon as the enclave receives the confirmation that the action performed in step 3 was done successfully, it issues a transaction to the blockchain that transfers the reward to the relevant \user.

\subsection{\name{} System Details}
We now present the protocols and message flow within \name{}. To facilitate the analysis, we describe the protocol as if a single \enclinf is present. We discuss in Section~\ref{sec:decentralization} how this design can be distributed to support multiple \enclinfs.

The \name{} protocol consists of three main parts, depicted in the right-hand side of Figure~\ref{fig:system_protocol}: enrollment of \users, campaign creation, and automatic interactions. We outline each one of them in the following sections.

\subsubsection{Enrollment of \UsErs}
\label{sec:user_enrollment}

Three prerequisites have to be met by a user to enroll in \name{} as an \user: the user needs to have credentials for a valid account for a service supported by the \name{} enclave, a cryptocurrency address to receive rewards, and a proxy on one of its devices.

The \name{} enclave exposes a web interface in which the \users can create a \name{} account, where all the user's information can be entered. The connection is secured with a TLS connection whose endpoints are the \user's device and the enclave, such that not even the OS in which the enclave is running can observe the data exchanged by the \user. The code of the enclave is public, and as part of the connection establishment, the \user checks the attestation of the enclave to make sure that it is communicating with a legitimate \name enclave running inside a genuine SGX CPU. %

Once the \user has established a secure connection with the \name{} enclave, it provides the credentials of its proxy server (if any are specified). Before proceeding further, the \name{} enclave ensures that the proxy is working correctly by sending a nonce to itself trough the user's proxy. As soon as the proxy is operational, the user can start providing the credentials and policies of the various accounts it wishes to rent out.

The enclave provides a list of services to the \user, for example various social media platforms. For each service, the \user inputs his own credentials, and selects a set of policies. The policies are specific to each service, and can restrict the type of content that can be associated with the \users through \name{}, or what kind of actions will be performed on their behalf. For instance, policies can be set to allow only ``likes'', or only pictures to be posted on a social network, or, in an e-voting platform, vote for candidates only if they belong to a particular political party. The enclave then checks the credentials of the service by trying to log-in to each one of them. After a successful login the service is considered as successfully enrolled.

The \user finally provides the cryptocurrency address to receive rewards. From this point on the various accounts entered by the \user can be rented by \requesters.

\subsubsection{Campaign Creation}
\label{sec:campaign_creation}
The \name{} enclave exposes a separate web interface to \requesters. The connection to this web interface is set up in the same way as described in Section~\ref{sec:user_enrollment} for the \users. That is, the TLS endpoints are at the \requesters browsers and in the enclave, and before exchanging any secret the \requester attests to the code of the \name enclave. 

As opposed to \users (cf.\ Section~\ref{sec:user_enrollment}), \requesters do not to create any account with the \name enclave and do not need to have any proxy installed. \Requesters are provided with a set of supported automatic actions for each service. As an example, for a social media campaign, they can provide a link to a post and the number of likes they wish to reach on that post, or they can provide a user and the number of new ``followers'' they want that user to obtain; in an e-voting service, \requesters can provide the name of a candidate and the number of votes they want it to receive.

Five main steps have to be completed to start a campaign. First, the \requester fills in the campaign details. Second, the enclave displays the expected price\footnote{The enclave always provides an upper bound, since the actual price is determined by the individual price of each account used.} for the specified campaign. Third, the \requester covers the amount of the campaign and an upfront deposit by transferring money to the cryptocurrency address of the \name enclave. The deposit will be refunded to the \requester during the campaign, and we use it to guarantee fairness, as explained in Section~\ref{sec:security_analysis}. Fourth, the \requester provides its cryptocurrency address, a transaction ID, and the latest block of the blockchain to the enclave. Finally, once the \name enclave verifies that the transaction exists, it has enough confirmation blocks and that the amount transferred is sufficient, it starts the campaign.

\subsubsection{Automatic Interactions}
\label{sec:automatic_interactions}
As soon as the campaign begins the enclave starts contacting \users whose policies are compatible with the actions of the campaign, we call the accounts of the \users that are compatible with the campaign ``\emph{compliant accounts}''. For each \user that has a compliant account, the enclave performs two main operations. First, it uses the proxy in the \user's enrolled device to get the latest block from the point of view of that \user. If the retrieved block is \emph{not} consistent (cf.\ Appendix~\ref{sec:back_crypto}) with the one sent by the \requester at campaign creation, the \name enclave does not perform any action for this \user in this campaign.
Second, if the previous step was successful, it connects to the service through the \user's proxy. By doing so, from the point of view of the service the connection appears as if it is initiated from the \user's device. Once the \name{} enclave establishes the connection with the service, it performs the action requested in the campaign on behalf of the \user. The enclave waits for a confirmation from the service. For some types of actions, the \requesters might require that the action is not manually reverted by the \users within a specified time-frame, in these cases the enclave delays the check until the end of the defined time-frame. Upon successful verification, the enclave issues a transaction on the blockchain that pays the \user in the address specified during enrollment and returns a share of the deposit to the \requester.

When the target number of actions of the campaign is reached, or if there are no more compliant accounts to employ, the enclave terminates the campaign. As part of the termination, any remaining campaign funds and the rest of the deposit are returned to the \requester on the cryptocurrency address specified at campaign creation.

\subsection{Protocol}

\begin{figure}[tbp]
    \centering
    \resizebox{\columnwidth}{!}{
        \tikzstyle{rect}=[rectangle,draw=black!50,thick,fill=white,align=center,minimum height=0.75cm,minimum width=2cm]

\tikzstyle{rectnone}=[rectangle,draw=none,thick,fill=white,align=center,minimum height=2.2cm,minimum width=1.8cm]
\tikzstyle{txt}=[rectangle,draw=white,fill=white,align=center,minimum width=2cm]

\tikzstyle{icon}=[rectangle,draw=white!50,fill=white,align=center,minimum height=1cm,minimum width=2cm]

\tikzset{
	partial ellipse/.style args={#1:#2:#3}{
		insert path={+ (#1:#3) arc (#1:#2:#3)}
	}
}

\tikzset{
	ncbar angle/.initial=90,
	ncbar/.style={
		to path=(\tikztostart)
		-- ($(\tikztostart)!#1!\pgfkeysvalueof{/tikz/ncbar angle}:(\tikztotarget)$)
		-- ($(\tikztotarget)!($(\tikztostart)!#1!\pgfkeysvalueof{/tikz/ncbar angle}:(\tikztotarget)$)!\pgfkeysvalueof{/tikz/ncbar angle}:(\tikztostart)$)
		-- (\tikztotarget)
	},
	ncbar/.default=0.5cm,
}

\begin{tikzpicture}[every node/.style={transform shape},apply style/.code={\tikzset{#1}},>=stealth']
\def\dist{0.7cm}
\def\hordistsmall{0.8cm}
\def\hordist{1.6cm}
\def\hordisttot{3cm}
\def\slope{0.3}
\def\spacing{0}

\node[icon] (blockchain) {\pgfuseimage{blockchain}};
\node[txt,above=0cm of blockchain] (t1) {Cryptocurrency};
\node[right=\hordistsmall of blockchain,icon] (buyer) {\pgfuseimage{renter}};
\node[txt] at (buyer |- t1) {\Requester};
\node[right=\hordist of buyer,icon] (enclave) {\pgfuseimage{enclave}} {};
\node[txt] (enclave-text) at (enclave |- t1) {Enclave};
\node[right=\hordist of enclave,icon] (user) {\pgfuseimage{owner}};
\node[txt] at (user |- t1) {\User};
\node[right=\hordistsmall of user,icon] (service) {\pgfuseimage{server}};
\node[txt] at (service |- t1) {Service};

\def\enroll{2.2}
\def\campaign{6.3}
\def\action{12.2}

\node[draw=none,minimum width=2.6cm] (enclave-end) [below=\action*\dist+2.9*\dist of enclave] {};
\node[draw=none] [above=0.1cm of enclave-text] {OS};
\node[draw=gray!50,line width=1mm,fit={(enclave) (enclave-end) (enclave-text)}] {};

\draw[-,thick] ($(blockchain)-(0, 1.2)$) -- ($(blockchain)-(0,\action*\dist + 4.7*\dist)$);
\draw[-,thick] ($(buyer)-(0, 1.2)$) -- ($(buyer)-(0,\action*\dist + 4.7*\dist)$);
\draw[-,thick] ($(enclave)-(0, 1.2)$) -- ($(enclave)-(0,\action*\dist + 4.7*\dist)$);
\draw[-,thick] ($(user)-(0, 1.2)$) -- ($(user)-(0,\action*\dist + 4.7*\dist)$);
\draw[-,thick] ($(service)-(0, 1.2)$) -- ($(service)-(0,\action*\dist + 4.7*\dist)$);

\draw[-,fill=white] ($(user)-(0, \action*\dist - 0.3*\dist)$) [partial ellipse=90:270:0.1cm and 0.15cm];
\draw[-,fill=white] ($(user)-(0, \action*\dist + 1.5*\dist)$) [partial ellipse=90:270:0.1cm and 0.15cm];
\draw[-,fill=white] ($(user)-(0, \action*\dist + 2.8*\dist)$) [partial ellipse=90:270:0.1cm and 0.15cm];

\foreach \i/\s/\l/\r/\t/\o/\h/\v/\c in {
	2.6/<-/enclave/user/Payment Address/-1/0.9/0.21/black!100,
	1.8/<-/enclave/user/Proxy Settings/-1/0.8/0.21/black!100,
	1.0/<-/enclave/user/Credentials/-1/0.9/0.14/black!100,
	0.3/<-/enclave/user/Attestation/-1/0.9/0.14/black!100
} {
	\draw[thick,sloped,apply style/.expand once=\s,color=\c] ($(\l)-(-\spacing,\i*\dist+\enroll*\dist)$) -- node[above,text width=\h*\hordisttot,align=center,fill=white,yshift=\v cm,anchor=base]{\t} ($(\r) -(\spacing,\i*\dist+\enroll*\dist + \o*\slope*\dist)$);	
}

\foreach \i/\s/\l/\r/\t/\o/\h/\v/\c in {
	-0.6/->/buyer/enclave/Attestation/+1/0.9/0.14/black!100,
	0.3/->/buyer/enclave/Set Options/+1/0.9/0.21/black!100,
	1.2/<-/blockchain/buyer/Deposit Funds/-1/0.8/0.21/black!100,
	2.1/<-/blockchain/buyer/Deposit Escrow/-1/0.8/0.21/black!100,
	2.6/->/buyer/enclave/Campaign/1/0.9/0.21/black!100,
	3.4/->/buyer/enclave/Latest Block/+1/0.9/0.14/black!100
} {
	\draw[thick,sloped,apply style/.expand once=\s,color=\c] ($(\l)-(-\spacing,\i*\dist+\campaign*\dist)$) -- node[above,text width=\h*\hordisttot,align=center,fill=white,yshift=\v cm,anchor=base]{\t} ($(\r) -(\spacing,\i*\dist+\campaign*\dist + \o*\slope*\dist)$);	
}

\foreach \i/\s/\l/\r/\t/\o/\h/\v/\f/\c in {
	-0.9/<-/enclave/user/Poll/-1/0.5/0.14/white/gray!100,
	0/<-/enclave/user/Latest Block/-1/0.9/0.14/white/black!100,
	1.2/->/enclave/service/Perform Action/+2/1/0.29/white/black!100,
	3.1/<-/enclave/service/Response/-2/1/0.34/white/black!100,
	3.8/->/enclave/user/Reward TX/1/0.7/0.14/white/black!100,
	4.1/<-/buyer/enclave/Return Deposit/-1/0.9/0.21/white/black!100
} {
	\draw[thick,sloped,apply style/.expand once=\s,color=\c] ($(\l)-(-\spacing,\i*\dist + \action*\dist)$) -- node[above,text width=\h*\hordisttot,align=center,fill=\f,yshift=\v cm,anchor=base]{\t} ($(\r) -(\spacing,\i*\dist + \action*\dist + \o*\slope*\dist)$);	
}

\draw[thick,<-] ($(blockchain) - (\spacing,\action*\dist - 1.2*\dist)$) to [ncbar=-0.3cm] ($(blockchain) - (\spacing,\action*\dist + 4*\dist)$);
\node[draw=none,align=center,color=black!100,rotate=90] at ($(blockchain) - (\spacing+0.5cm,\action*\dist + 1.5*\dist)$) {Repeat};

\draw[thick,->,color=black!100] ($(enclave) - (\spacing,\action*\dist - 0*\dist)$) to [out=180,in=180,looseness=2] node[left,text width=\hordist*1.2,align=center,fill=white]{Check Consistency} ($(enclave) - (\spacing,\action*\dist + 1*\dist)$);

\draw[-,fill=white] ($(user)-(0.01, \action*\dist - 0.3*\dist)$) [partial ellipse=90:-90:0.1cm and 0.15cm];
\draw[-,fill=white] ($(user)-(0.01, \action*\dist + 1.5*\dist)$) [partial ellipse=90:-90:0.1cm and 0.15cm];
\draw[-,fill=white] ($(user)-(0.01, \action*\dist + 2.8*\dist)$) [partial ellipse=90:-90:0.1cm and 0.15cm];

\def\bspace{0.2}
\draw[thick,gray,dashed] ($(service)-(+0.1,\enroll*\dist+2.9*\dist)$) -- ($(blockchain)-(-0.1,\enroll*\dist+2.9*\dist)$);
\draw[thick,gray,decorate,decoration={brace,amplitude=6pt}] ($(service)-(-\bspace,\enroll*\dist-0.3*\dist)$) -- ($(service)-(-\bspace,\enroll*\dist+2.7*\dist)$);
\node[gray,rotate=90,below] at ($(service)-(-\bspace - 0.2,\enroll*\dist+1.1*\dist)$) {Enrollment};

\draw[thick,gray,decorate,decoration={brace,amplitude=6pt}] ($(service)-(-\bspace,\campaign*\dist-1.0*\dist)$) -- ($(service)-(-\bspace,\campaign*\dist+3.8*\dist)$);
\node[gray,rotate=90,below] at ($(service)-(-\bspace-0.2,\campaign*\dist+1.4*\dist)$) {Campaign Creation};

\draw[thick,gray,dashed] ($(service)-(+0.1,\campaign*\dist+4.0*\dist)$) -- ($(blockchain)-(-0.1,\campaign*\dist+4.0*\dist)$);
\draw[thick,gray,decorate,decoration={brace,amplitude=6pt}] ($(service)-(-\bspace,\action*\dist-2*\dist)$) -- ($(service)-(-\bspace,\action*\dist+4.1*\dist)$);
\node[gray,rotate=90,below] at ($(service)-(-\bspace - 0.2,\action*\dist+1.05*\dist)$) {Automatic interactions};

\draw[-,thick] ($(enclave)-(0.5cm, \enroll*\dist+1.4*\dist)$) circle (0.25cm) node {\textbf{1}};
\draw[-,thick] ($(buyer)-(0.5cm, \campaign*\dist-0.5*\dist)$) circle (0.25cm) node {\textbf{2}};
\draw[-,thick] ($(buyer)-(-0.5cm, \campaign*\dist+1.2*\dist)$) circle (0.25cm) node {\textbf{3}};
\draw[-,thick] ($(enclave)-(-0.5cm, \campaign*\dist+3.2*\dist)$) circle (0.25cm) node {\textbf{4}};
\draw[-,thick] ($(enclave)-(0.5cm, \action*\dist-0.7*\dist)$) circle (0.25cm) node {\textbf{5}};
\draw[-,thick] ($(user)-(-1.0cm, \action*\dist+2.15*\dist)$) circle (0.25cm) node {\textbf{6}};
\draw[-,thick] ($(enclave)-(0.5cm, \action*\dist+4.45*\dist)$) circle (0.25cm) node {\textbf{7}};

\end{tikzpicture}
    }
    \caption{The protocol of \name{}.}
    \label{fig:system_protocol}
\end{figure}
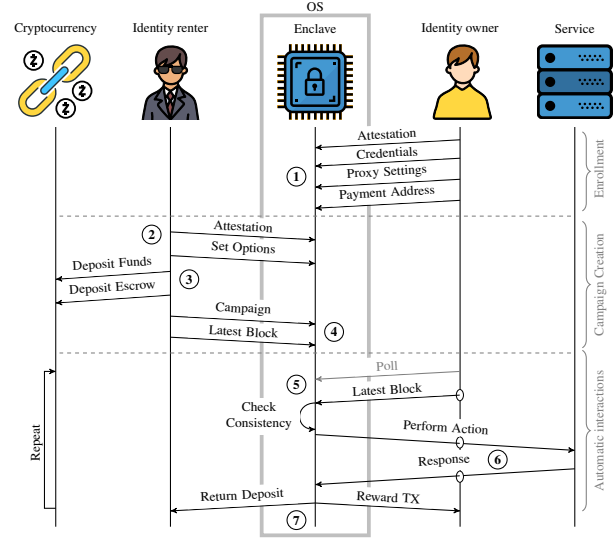

We now put together all the steps of the protocol described in the previous subsections. We depict the general flow of a \name campaign in Figure~\ref{fig:system_protocol}, where each number in the figure corresponds to the following steps:
\begin{enumerate}
    \item \Users enroll in \name. They attest the enclave and provide the enclave access to a proxy running on their devices, their credentials for various services, and a set of policies for these services.
    \item An \requester connects to the enclave and specifies the details of a campaign. %
    \item The \requester provides the funds necessary for the campaign and a deposit to the \name enclave's cryptocurrency address.
    \item The \requester sends the final campaign confirmation and the latest block to the enclave. The enclave then checks if the funds are received correctly and selects compliant accounts to fulfill the campaign. %
    \item The enclave fetches the latest block through each compliant accounts proxy and checks the consistency with the block provided by the \requester.
    \item The enclave performs the action via the proxy on the \user's device.
    \item The enclave issues the reward transaction to the respective compliant account. A share of the deposit is immediately returned to the \requester. If specified by the \requester, the enclave waits a specified time and checks that the action has not been reverted before issuing the transaction.
\end{enumerate}

Once the campaign is over, or if no more compliant accounts can be reached, the enclave returns the remaining campaign funds and deposit to the \requester.

\subsection{The \name{} Enclaves}
\label{sec:mult_encl}
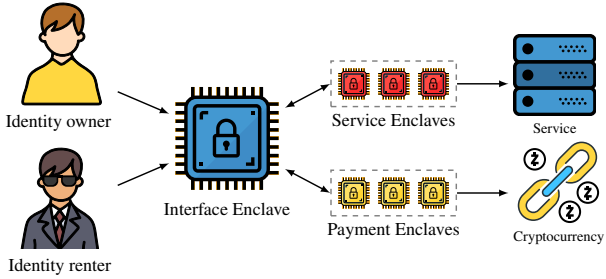
\begin{figure}[tbp]
    \centering
    \resizebox{\columnwidth}{!}{
        \tikzstyle{place}=[rectangle,draw=black!50,thick,fill=white,align=center,minimum height=0.8cm,minimum width={{width("Twitter")+15pt}}]
\tikzstyle{icon}=[rectangle,draw=white!50,fill=white,align=center,minimum height=1cm,minimum width=2cm]

\tikzset{every loop/.style={min distance=10mm,in=240,out=300,looseness=5}}
\begin{tikzpicture}
\node[icon] (E)                    {\pgfuseimage{enclave}};
\node[draw=none] [below=0cm of E] {Interface Enclave};
\node[icon] (U) [above left=-1.1cm and 1cm of E]  {\pgfuseimage{owner}\\\User};
\node[icon] (B) [below left=-1.1cm and 1cm of E]  {\pgfuseimage{renter}\\\Requester};

\node[icon,scale=0.3] (S1) [above right=-0.4cm and 1cm of E] {\pgfuseimage{enclave-yellow}};
\node[icon,scale=0.3] (S2) [right=0.1cm of S1] {\pgfuseimage{enclave-yellow}};
\node[icon,scale=0.3] (S3) [right=0.1cm of S2] {\pgfuseimage{enclave-yellow}};
\node[draw=black!50,dashed,fit={(S1) (S2) (S3)}, inner sep=3pt, thick] (S) {};
\node[draw=none] [below=0cm of S] {Service Enclaves};

\node[icon,scale=0.3] (Z1) [below right=-0.4cm and 1cm of E] {\pgfuseimage{enclave-red}};
\node[icon,scale=0.3] (Z2) [right=0.1cm of Z1] {\pgfuseimage{enclave-red}};
\node[icon,scale=0.3] (Z3) [right=0.1cm of Z2] {\pgfuseimage{enclave-red}};
\node[draw=black!50,dashed,fit={(Z1) (Z2) (Z3)}, inner sep=3pt,thick] (Z) {};
\node[draw=none] [below=0cm of Z] {Payment Enclaves};

\node[icon, scale=0.8] (ZCash) [right=of Z] {\pgfuseimage{blockchain}\\Cryptocurrency};
\node[icon, scale=0.8] (Service) [right=of S] {\pgfuseimage{server}\\Service};

\path[->,thick]   (U) edge (E);
\path[->,thick]   (B) edge (E);
\path[<->,thick]  (E) edge ( $(S.west)$ );
\path[<->,thick]  (E) edge ( $(Z.west)$ );
\path[->,thick]   (Z) edge (ZCash);
\path[->,thick]   (S) edge (Service);

\end{tikzpicture}
    }
    \caption{Architecture of the \name enclaves.}
    \label{fig:enclave}
\end{figure}

So far we considered the \name{} enclave as a single enclave that takes care of all the interactions between all the entities in \name. However, in the interest of modularity (and later scalability), we split the enclave into three parts: (1) the interface enclave, (2) the service enclave, and (3) the payment enclave. We depict them and how they interact with each other in Figure~\ref{fig:enclave}. Note that there is an enlistment phase in which the enclaves attest each other, and at this time they exchange their public keys to later establish a secure TLS channel between them without having to repeat attestation.

\paragraph{Interface Enclave.}
The interface enclave is the entry point to all the other enclaves: Users of \name{} (\users and \requesters) will use it to start campaigns and enroll their credentials. The interface enclave stores all enrolled credentials and synchronizes the actions of the service enclaves and the payment enclaves. During an enlistment phase the interface enclaves allows an administrator to enroll service and payment enclaves. %

\paragraph{Service Enclave.}
A different service enclave exists for each supported service. For instance, to support a particular social network, a service enclave that knows how to interface with that social network needs to be developed. The service enclave exposes a list of \emph{service specific} tasks and policies to the interface enclave. The attestation report of all service enclaves is also exposed by the interface enclave, so that users can decide to trust only a particular implementation of a service enclave.

The service enclave receives proxy configuration parameters, credentials, and policies from the interface enclave and tries to fulfill the received tasks by contacting the service through the \user's proxy. It then sends confirmations of completed actions back to the interface enclave. The service enclaves do not keep any permanent information about the \users data. Each service enclave is independent of each other, and allows to scale the number of requests made to the service, as the more enclaves there are the more tasks can be performed in parallel.

\paragraph{Payment Enclave.}
\label{sec:protocol:payment}
Similar to the service enclaves there can be multiple implementations of payment enclaves, in this case, each supporting a different digital currency (for more details see Appendix~\ref{sec:background}). However, to ease explanation, we assume there exists a single implementation that supports ZCash. 

At the beginning of the campaign, funds are split into smaller shares each of which is controlled by one payment enclave which can then issue transactions independently. %
To guarantee that all the reward transactions are correctly issued, all transactions of one payment enclave depend on each other: each following transaction uses the unspent output of the previous transaction. Since  a share of the deposit is released with every reward transaction, we can guarantee that if the entire deposit is returned then all rewards have been paid as well. 

The payment enclaves share a backup copy of their private keys to access the funds with the interface enclave. In the case of a crash of a payment enclave, the interface enclave can use said keys to restore the funds.

\section{Security Analysis}
\label{sec:security_analysis}
In this section we informally analyze the security of \name through provided guarantees of the three main system properties -- fairness, indistiguishability, and plausible deniability (cf.\ Section~\ref{sec:prob_statement}). We look at two different adversary models. First, we consider the case in which every protocol participant and third parties try to violate fairness. Second, for the latter two properties, we consider a powerful attacker that wants to expose the participants. Thus, the various protocol participants can cooperate to prevent disruption of \name.

Regarding \name enclaves, we assume the standard SGX adversary model: the adversary controls the OS on the platform where the enclaves run, and can tap into the memory bus, but cannot tamper with the CPU package. Additionally, the adversary can delay or drop any network packet -- however, she cannot see or modify the content of packets if the enclave serves as the TLS endpoint.
We assume no adversary can compromise the enclaves and we trust the TEE manufacturer (i.e., Intel) to not fake attestation queries, or otherwise compromise the security of any SGX capable CPUs, either voluntarily or under coercion from an external party. Rollback attacks are out of scope -- we refer the interested reader to~\cite{matetic2017rote,brandenburger2017rollback}. 

\subsection{Violating Fairness}
\label{sec:fairness}
Parties participating in the protocol are interested in violating fairness to get economical advantages. However, external entities (e.g., services targeted by \name) can also be interested in violating fairness in the hopes of damaging the reputation of \name and push \users and \requesters to leave the platform. We first detail the considered threat model, then analyze security against protocol participants, and finally against external adversaries.

\paragraph{Threat Model.}
All three protocol parties want to break fairness, i.e., the \requester wants to obtain actions without paying for them, the \user wants to get rewards without performing actions, and the \enclinf wants to get fees without doing the required work. Parties might also collude to break fairness. We assume that protocol parties act rationally and are moved by economic reasons, thus would not do any action that could impact their stakes in the system. However, there might be external adversaries compromising any of the parties -- such entities can act without the fear of economic repercussions  and can make the compromised parties act against their interest. We do not provide any fairness guarantee to these compromised parties, and analyze how they would impact the remaining honest parties.

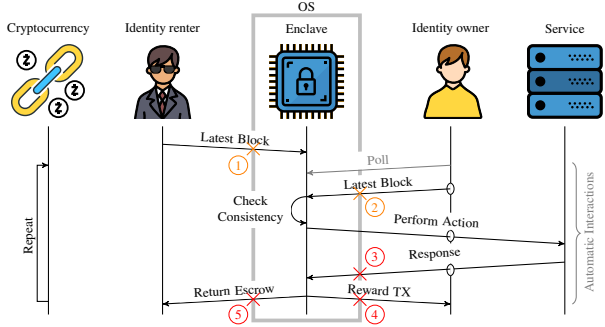
\begin{figure}[t]
    \centering
    \resizebox{\columnwidth}{!}{
    \tikzstyle{rect}=[rectangle,draw=black!50,thick,fill=white,align=center,minimum height=0.75cm,minimum width=2cm]

\tikzstyle{rectnone}=[rectangle,draw=none,thick,fill=white,align=center,minimum height=2.2cm,minimum width=1.8cm]
\tikzstyle{txt}=[rectangle,draw=white,fill=white,align=center,minimum width=2cm]

\tikzstyle{icon}=[rectangle,draw=white!50,fill=white,align=center,minimum height=1cm,minimum width=2cm]

\tikzset{
	partial ellipse/.style args={#1:#2:#3}{
		insert path={+ (#1:#3) arc (#1:#2:#3)}
	}
}

\tikzset{
	ncbar angle/.initial=90,
	ncbar/.style={
		to path=(\tikztostart)
		-- ($(\tikztostart)!#1!\pgfkeysvalueof{/tikz/ncbar angle}:(\tikztotarget)$)
		-- ($(\tikztotarget)!($(\tikztostart)!#1!\pgfkeysvalueof{/tikz/ncbar angle}:(\tikztotarget)$)!\pgfkeysvalueof{/tikz/ncbar angle}:(\tikztostart)$)
		-- (\tikztotarget)
	},
	ncbar/.default=0.5cm,
}

\tikzset{cross/.style={cross out, draw=black, minimum size=2*(#1-\pgflinewidth), inner sep=0pt, outer sep=0pt,thick},cross/.default={0.2cm}}

\begin{tikzpicture}[every node/.style={transform shape},apply style/.code={\tikzset{#1}},>=stealth']
\def\dist{0.7cm}
\def\hordistsmall{0.8cm}
\def\hordist{1.6cm}
\def\hordisttot{3cm}
\def\slope{0.3}
\def\spacing{0}

\node[icon] (blockchain) {\pgfuseimage{blockchain}};
\node[txt,above=0cm of blockchain] (t1) {Cryptocurrency};
\node[right=\hordistsmall of blockchain,icon] (buyer) {\pgfuseimage{renter}};
\node[txt] at (buyer |- t1) {\Requester};
\node[right=\hordist of buyer,icon] (enclave) {\pgfuseimage{enclave}} {};
\node[txt] (enclave-text) at (enclave |- t1) {Enclave};
\node[right=\hordist of enclave,icon] (user) {\pgfuseimage{owner}};
\node[txt] at (user |- t1) {\User};
\node[right=\hordistsmall of user,icon] (service) {\pgfuseimage{server}};
\node[txt] at (service |- t1) {Service};

\def\enroll{0}
\def\campaign{0}
\def\action{4.4}

\node[draw=none,minimum width=2.6cm] (enclave-end) [below=\action*\dist+2.6*\dist of enclave] {};
\node[draw=none] [above=0.1cm of enclave-text] {OS};
\node[draw=gray!50,line width=1mm,fit={(enclave) (enclave-end) (enclave-text)}] {};

\draw[-,thick] ($(blockchain)-(0, 1.2)$) -- ($(blockchain)-(0,\action*\dist + 4.5*\dist)$);
\draw[-,thick] ($(buyer)-(0, 1.2)$) -- ($(buyer)-(0,\action*\dist + 4.5*\dist)$);
\draw[-,thick] ($(enclave)-(0, 1.2)$) -- ($(enclave)-(0,\action*\dist + 4.5*\dist)$);
\draw[-,thick] ($(user)-(0, 1.2)$) -- ($(user)-(0,\action*\dist + 4.5*\dist)$);
\draw[-,thick] ($(service)-(0, 1.2)$) -- ($(service)-(0,\action*\dist + 4.5*\dist)$);

\draw[-,fill=white] ($(user)-(0, \action*\dist - 0.3*\dist)$) [partial ellipse=90:270:0.1cm and 0.15cm];
\draw[-,fill=white] ($(user)-(0, \action*\dist + 1.5*\dist)$) [partial ellipse=90:270:0.1cm and 0.15cm];
\draw[-,fill=white] ($(user)-(0, \action*\dist + 2.8*\dist)$) [partial ellipse=90:270:0.1cm and 0.15cm];

\foreach \i/\s/\l/\r/\t/\o/\h/\v/\f/\c in {
	-2/->/buyer/enclave/Latest Block/+1/0.9/0.14/white/black!100,
	-0.9/<-/enclave/user/Poll/-1/0.5/0.14/white/gray!100,
	0/<-/enclave/user/Latest Block/-1/0.9/0.14/white/black!100,
	1.2/->/enclave/service/Perform Action/+2/1/0.29/white/black!100,
	3.1/<-/enclave/service/Response/-2/1/0.34/white/black!100,
	3.8/->/enclave/user/Reward TX/1/0.7/0.14/white/black!100,
	4.1/<-/buyer/enclave/Return Escrow/-1/0.9/0.14/white/black!100
} {
	\draw[thick,sloped,apply style/.expand once=\s,color=\c] ($(\l)-(-\spacing,\i*\dist + \action*\dist)$) -- node[above,text width=\h*\hordisttot,align=center,fill=\f,yshift=\v cm,anchor=base]{\t} ($(\r) -(\spacing,\i*\dist + \action*\dist + \o*\slope*\dist)$);	
}

\draw[thick,<-] ($(blockchain) - (\spacing,\action*\dist - 1.2*\dist)$) to [ncbar=-0.3cm] ($(blockchain) - (\spacing,\action*\dist + 4*\dist)$);
\node[draw=none,align=center,color=black!100,rotate=90] at ($(blockchain) - (\spacing+0.5cm,\action*\dist + 1.5*\dist)$) {Repeat};

\draw[thick,->,color=black!100] ($(enclave) - (\spacing,\action*\dist - 0*\dist)$) to [out=180,in=180,looseness=2] node[left,text width=\hordist*1.2,align=center,fill=white]{Check Consistency} ($(enclave) - (\spacing,\action*\dist + 1*\dist)$);

\draw[-,fill=white] ($(user)-(0.01, \action*\dist - 0.3*\dist)$) [partial ellipse=90:-90:0.1cm and 0.15cm];
\draw[-,fill=white] ($(user)-(0.01, \action*\dist + 1.5*\dist)$) [partial ellipse=90:-90:0.1cm and 0.15cm];
\draw[-,fill=white] ($(user)-(0.01, \action*\dist + 2.8*\dist)$) [partial ellipse=90:-90:0.1cm and 0.15cm];

\def\bspace{0.2}

\draw[thick,gray,decorate,decoration={brace,amplitude=6pt}] ($(service)-(-\bspace,\action*\dist-1.3*\dist)$) -- ($(service)-(-\bspace,\action*\dist+4.1*\dist)$);
\node[gray,rotate=90,below] at ($(service)-(-\bspace - 0.2,\action*\dist+1.35*\dist)$) {Automatic Interactions};

\def\halfdist{1.425cm}
\draw[-,color=orange] ($(enclave)-(\halfdist + 0.4cm, \action*\dist-1.2*\dist)$) circle (0.25cm) node {1};
\draw ($(enclave)-(\halfdist, \action*\dist-1.83*\dist)$) node[cross,orange] {};
\draw[-,color=orange] ($(enclave)-(-\halfdist - 0.4cm, \action*\dist+0.4*\dist)$) circle (0.25cm) node {2};
\draw ($(enclave)-(-\halfdist, \action*\dist-0.125*\dist)$) node[cross,orange] {};
\draw[-,color=red] ($(enclave)-(-\halfdist-0.4cm, \action*\dist+2.3*\dist)$) circle (0.25cm) node {3};
\draw ($(enclave)-(-\halfdist, \action*\dist+2.95*\dist)$) node[cross,red] {};
\draw[-,color=red] ($(enclave)-(-\halfdist-0.4cm, \action*\dist+4.5*\dist)$) circle (0.25cm) node {4};
\draw ($(enclave)-(-\halfdist, \action*\dist+3.92*\dist)$) node[cross,red] {};
\draw[-,color=red] ($(enclave)-(\halfdist+0.4cm, \action*\dist+4.5*\dist)$) circle (0.25cm) node {5};
\draw ($(enclave)-(\halfdist, \action*\dist+3.92*\dist)$) node[cross,red] {};
\end{tikzpicture}
    }
    \caption{Adversaries controlling the OS of the \enclinf, or the local network of protocol participants, can cut network connections at the points marked 1--5 to try to disrupt the protocol and its properties. Note that the content of message 1 and 2 can be modified by the \requesters and \users, respectively, since there is no TLS connection to the blockchain nodes.}
    \label{fig:security_protocol}
\end{figure}

Given adversarial capabilities and the assumed SGX adversary model, we highlight that no misbehaving party can impact the protocol by modifying execution of the \name enclave at any step, or by modifying the content of network messages. Hence, an adversary can only affect fairness by cutting network messages of the protocol: Figure~\ref{fig:security_protocol} depicts all susceptible messages.

\paragraph{Protocol Participants.}
We now analyze the fairness guarantees of \name when only protocol participants try to cut connections and messages.
The \requester initiates a campaign by sending its current view of the blockchain to the enclave. The \requester has full control over the content of this message, and could send the \name enclave a different view of the blockchain if he so wishes (message (1) in Figure~\ref{fig:security_protocol}). Then, by cutting message (2) of Figure~\ref{fig:security_protocol}, the \requester could try to trick the enclave into performing an action without actually having the funds. The attacker hence effectively forges the enclave's view of the blockchain. However, \name{} will only perform an action on behalf of an \user if and only if the blockchain state from the \user is consistent with the one provided by the \requester.
If an adversary can carry out a successful double spend attack against some \users, then he can break the fairness property for these specific parties, but not against other unaffected \users.

The \requester might also try to cut the response of the external service, that is message (3) in Figure~\ref{fig:security_protocol}. The enclave will then never get a confirmation of the completed action but it will also not get an error message, and it is, therefore, able to distinguish an attack from incorrect credentials\footnote{Note that \users could try to fake message (3) by exposing a controlled copy of the target service to the enclave -- however, this can be prevented by existing countermeasures such as certificate pinning~\cite{evans2015public}.}. In this case, the \requester would lose a part of his deposit. Messages (4) and (5) in Figure~\ref{fig:security_protocol} are intrinsically linked since both the reward and the share of the deposit are linked into a single transaction. By blocking this transaction, the \requester would lose his share of the deposit. 

Any of the previously mentioned messages could also be blocked by the \enclinf. However, he would forfeit his fees, since the \enclinf gets the fees with the reward transaction (messages (4) and (5) in Figure~\ref{fig:security_protocol}), so he has incentives to keep the system running as long as possible.

In conclusion, tampering with any of the five messages highlighted in Figure~\ref{fig:security_protocol} would result in losses for all parties (even when colluding): \requesters (escrow), \users (reward) and \enclinfs (service fees) -- thus, no rational party would perform this attack.

\paragraph{External Adversaries.}
As discussed, an external adversary can compromise any protocol party and make them act irrationally to try to break fairness for other parties (no guarantee of fairness is given to the compromised party). We now analyze what happens to the remaining protocol participants:

\begin{itemize}
    \item A compromised \textbf{\requester} cannot impact the fairness for any other party, since it can only tamper with messages (1) and (4) of Figure~\ref{sec:security_analysis}. As discussed, tampering with message (1) has no implication for \users that have the correct view of the blockchain, and cutting message (4) does not prevent the reward to the \user because she receives a copy of this transaction as well in message (5).
    \item A compromised \textbf{\user} might cut message (3) and forfeit the reward in order to destroy a small share of the deposit, thus hurting the \requester.
    \item A compromised \textbf{\enclinf} could either shut down the enclave upon receiving the funds, or he could cut messages (3) or the reward transaction in Figure~\ref{fig:security_protocol}, thus executing all actions but not paying any reward and destroying the deposit.
\end{itemize}

Additionally, a service targeted by \name is in the unique position of being able to fake the proof of performed action (message 3 of Figure~\ref{fig:security_protocol}). This can happen if the service colludes with \users (for example by registering a large set of fake \users to a \name enclave), therefore it knows when it receives a rented action and provides fake responses accordingly. If the result of the rented action is observable, such as a ``like'' on a OSN, this is trivially prevented: the enclave can verify ``externally'' (e.g., through a legit account) if the action was performed. However, if the result of the action is not observable, like in e-voting, the service can violate fairness successfully. We discuss a Peer-to-Peer design for \name that can mitigate this scenario by requiring one SGX CPU per enrolled account in Section~\ref{sec:decentralization}.

\subsection{Exposing Parties}
\label{sec:exposing_parties}
Services targeted by \name might try to limit its impact by breaking \emph{indistinguishability} and filtering fraudulent actions. In the case of e-voting, the service provider (i.e., a government) might also try to punish participants of \name with legal repercussions by breaking \emph{plausible deniability}. Here we first elaborate on such a threat model, then analyze the security against non colluding and colluding adversaries.

\paragraph{Threat Model.}
Indistinguishability and plausible deniability require a different threat model: uncompromised protocol participants might want to cheat on each other regarding fairness, but ultimately they still want to perform the exchange of money for identities. As they have monetary stakes in the system, they will not try to expose each other without any external influence, and can instead cooperate to avoid being exposed.
Therefore, the adversaries we consider here are external entities that can possibly compromise protocol participants (to expose other participants, as the exposure of the compromised party is trivial), and the services targeted by \name. All these adversaries can collude to reach their goal.

Once again, we assume the previously described standard SGX adversary model where the adversary can cut connections and power down enclaves, but the enclaves execution or the messages cannot be modified. Additionally, services have full control over their platform, and can run any kind of analysis on users' actions to try to find accounts that have been rented. As a discussion point, we also consider \textit{global} network attackers with a complete view of the network.

\paragraph{No Collusion.}
When assuming no collusion, the adversary is in full control of a single entity (e.g., the service) but cannot cooperate with any other entity. For \emph{indistinguishability}, assuming that protocol participants do not try to expose each other, the only possible adversary is the service. Indeed, services targeted by \name are the only ones that can analyze users' actions -- and thus can try to distinguish between fraudulent (i.e., rented) and genuine actions. 
To do so, services can analyze both users' behavior on their platforms, to understand if actions are performed automatically or by a human; services can also analyze the result of users' actions, e.g., posted or ``liked'' content, to try to detect fraudulent actions. These are active research areas, proposing techniques such as CAPTCHAs~\cite{von2003captcha}, or Machine Learning tools to build behavioral and content profiles to detect misbehaving users~\cite{gilani2017classification,viswanath2014towards} and their actions~\cite{jindal2008opinion,mukherjee2012spotting}. The literature on these topics is extensive -- we report the main research results in Section~\ref{sec:defences_existing_OSN}, where we discuss that there is still no conclusive countermeasure to stop systems such as \name. Here, we only highlight that such areas of detection are subject to cat-and-mouse games, where adversaries can often avoid detection by changing their strategy after defenders deploy their countermeasures~\cite{wang2014man,viswanath2014towards,ye2018yet,juuti2018stay}.

\paragraph{Colluding Adversary.}
A stronger adversary can collude with any party and, is possibly, a global network adversary. The service can collude with \requesters, and create campaigns with specific targets that therefore de-anonymize \users (e.g., a OSN asking to ``like'' a post that is only visible via a direct link, to prevent legitimate interactions). Another possible attack includes observing all the connections to and from any \name enclave, thus correlating them to the \user's proxy and to the service. Given the ability to then link these connections with the service's accounts it would be possible to identify \users participating in \name, breaking both indistinguishability and plausible deniability.
Even adding an anonymity network (e.g., Tor, as we discuss in Section~\ref{sec:adding_anon}) does not mitigate against this attacker, since a global network attacker is outside the Tor threat model.
Therefore under this attacker model, for services that allow to link actions with \users (e.g., social networks, but \emph{not} e-voting), \name cannot guarantee indistinguishability and plausible deniability.

\subsection{Anonymity}
\label{sec:anonymity}

In the general case, all operation on \name are anonymous: \requesters do not learn any detail about \users, whose credentials are securely stored and not accessible even by the \enclinfs. However, there could be specific actions and target services that can violate anonymity. For example, when the bought action has an observable result (e.g., interactions on OSNs, as opposed to e-voting), \users could see who is the target, and \requesters could try to correlate the start of campaigns and the results of the bought actions. Note that, even in this case, both parties can still appeal to plausible deniability.
More powerful adversaries are \enclinfs, who could track IP addresses of \users and \requesters; in general, powerful network adversaries could correlate IP addresses connecting to the enclave and subsequently to target services. If such a strong adversary would collude with the target service, they could directly expose users' rented accounts, and act accordingly. Network adversaries can be mitigated (with the exception of timing correlations) by using mix networks such as Tor~\cite{tor-design}, as we discuss in Section~\ref{sec:adding_anon}. While there is no general solution to deanonymization by specific actions, \users can specify policies regarding the actions that they are willing to sell, thus specifying the amount of risk they are willing to take.

\section{Defenses}
\label{sec:defences}
We now shift perspective and present some possible defenses against systems such as \name. These countermeasures extend our reasoning presented in Section~\ref{sec:security_analysis}, and follow from the most powerful adversary models we considered. We discuss existing countermeasures for the two applications we considered throughout this paper (OSNs and e-voting), and then elaborate on potential future measures. We attempt to order defenses by increased complexity for the defending entities, and highlight that all countermeasures require significant effort for the defenders. In particular, some countermeasures require an unrealistic amount of collaboration between nations and services while others only lessen the impact but do not entirely prevent the problem. In conclusion, we believe that none of these countermeasures is easily employable to defend against systems such as \name, and more research attention is necessary to stop this threat.

\subsection{Existing Techniques - OSNs}
\label{sec:defences_existing_OSN}
There already exist countermeasures against ``non-human'' accounts and interactions on OSNs, both deployed by commercial providers and proposed in the literature. We coarsely group them according to what OSNs can analyze: (i) users' behavior, trying to understand if the actions were performed automatically, and (ii) users' interactions, e.g., posted content, or targets of ``like'' actions, trying to uncover automatically generated content by \name. 

\paragraph{Users' Behavior.} 
The behaviour of authentic users can be fully imitated by \name. Detecting rented accounts is thus identical to the conventional bot detection problem.
We discuss two aspects of this problem and the relevant literature next.
First, services might try to use CAPTCHAs~\cite{von2003captcha} to detect automated actions. However, CAPTCHAs can be bypassed by advances in machine learning technologies~\cite{ye2018yet}, or by offloading them to cheap labor~\cite{motoyama2010re}. Services might also build users' behavioral profiles and try to detect ``compromised'' accounts; however, most approaches presented in literature are based on machine learning~\cite{adewole2017malicious} -- turning into a cat and mouse game where detection can be easily avoided by changing strategies~\cite{wang2014man,viswanath2014towards}.
Second, services might try to detect if users' behavior changes abruptly, or deviates for a long time from previous behavioral profilings~\cite{gilani2017classification,adewole2017malicious}. However, in a platform such as \name, rented actions are interleaved with \users' benign behavior, and it is thus unclear if any of these solutions applies to identity lease. OSNs can instead try to detect behavior similar to \name's \users: \textit{brief} deviations from normal behavior~\cite{viswanath2014towards}, and then try to correlate such deviations among multiple accounts~\cite{egele2013compa} in order to expose vast campaigns. Proposals in the area still rely on hand-crafted features that \name can manipulate, and plausible deniability still applies: \users can claim it was their legitimate interaction, therefore it is hard to argue that OSNs can easily start deleting posts and interactions.
Ultimately, we note that even though fake and compromised accounts are a well studied problem in research, OSNs still struggle to detect them~\cite{fb-fake-news,twitter2018midterms}, showing that this problem is far from trivial.

\paragraph{Users' Interactions.} 
Services are interested in trying to detect crowdturfed and fake content. There has been a lot of effort in detecting such content~\cite{jindal2008opinion,ott2012estimating}, accounts involved in crowdturfing activities~\cite{mukherjee2012spotting}, and even the targets of crowdturfing actions~\cite{song2015crowdtarget}. However, this is again a cat-and-mouse game, as recent efforts have shown how to evade such classifiers~\cite{wang2014man} and how to automatically generate realistic content that can fool even human readers~\cite{yao2017automated,juuti2018stay}. To conclude, some of the arguments for (i) still apply: plausible deniability is not violated, as there is no proof that \users did not generate the dubious content, and these countermeasures are even ineffective if there is no observable result.

\subsection{Existing Techniques - E-voting}
\label{sec:defences_existing_voting}
In this section we analyze how some security properties of e-voting schemes might defend against \name by preventing its operations, or by violating \name's requirements. In particular, here we look at three properties of e-voting and we refer the interested reader to Appendix~\ref{sec:evoting} for more details. First, verifiability, and in particular \emph{individual verifiability}, gives the voters confidence that their vote was recorded as intended. Second, coercion-resistance removes the possibility from voters to prove to third parties how (and whether) they voted. Finally, \emph{privacy} is a weaker form of coercion-resistance which prevents votes from being linked to a specific voter.

We premise our analysis by observing that, regarding interactions between an e-voting service and \name, the \name enclave acts as both the device with which the user votes, and as the voter itself, from the point of view of the voting service. This behaviour is referred to as a \textit{simulation attack}~\cite{juels2005coercion} in the literature since the enclave can perfectly emulate a voter. We note that voters successfully enroll as \users by providing all the secret that they needed to register and participate in the election.

\paragraph{Privacy.}
This property inadvertently helps the indistinguishability and plausible deniability properties of \name. Because of voter privacy, it is very challenging for the voting authorities to distinguish between voters giving their inputs to a legitimate voting device, and  the \name enclave autonomously voting on behalf of a voter --  because the \name enclave can be seen as a voting device.
Another consequence of voter privacy is that even in the extreme case in which the election authorities could tell which votes were cast with \name{}\footnote{For instance, because the election allows write-ins that were used only when voting with the \name enclave.}, they would not be able to link them to specific voters.

\paragraph{Verifiability and Coercion-Resistance.}
Verifiability~\cite{cortier2016sok, kusters2010accountability} and coercion-resistance~\cite{juels2005coercion, clark2011selections,clarkson2008civitas} have an impact on the fairness property of \name. In particular, the former aids it, while the latter could break it, thus potentially being a concrete defence for e-voting against \name.

A distinct difference from e-voting to OSNs is that in e-voting a \user can only sell one single action per election (i.e., one vote), as opposed to OSNs where usually actions can be performed and reverted an arbitrary number of times. Individual verifiability, in this case, helps to achieve fairness: since the \name enclave acts as the voter, it can check whether a vote has been cast and whether it reflects the choice that the \requester is willing to pay for.
This aspect is particularly relevant when one considers the different policy with regards to multiple ballots submissions for a voter. In particular, for each voter two policies are common: (i) only the first vote counts and all subsequent votes are discarded, (ii) only the last submitted ballot counts, and all previous ones are ignored. Individual verifiability allows to check whether any vote was submitted before (or whether no vote was submitted after, respectively) the \name enclave casts its vote.

Coercion-resistance~\cite{juels2005coercion, clark2011selections} is the most promising property to defend against \name. However, while coercion-resistance makes it impossible to prove to a third party that a voter voted in a particular way, the voter itself (in order to guarantee individual verifiability) has to be able to check the content of its own cast ballot. Therefore, \emph{if the \name enclave has valid voting credentials}, since it acts as a voter, coercion-resistance does little to prevent the enclave to verify that the vote was cast correctly - this is effectively a simulation attack. %
Hence, in order to prevent a simulation attack, in~\cite{juels2005coercion} it is proposed to allow the voter to generate arbitrary \emph{invalid} voting credentials which are indistinguishable from the real ones (the votes cast with these credentials are later discarded). However, since the initial credentials are delivered to the voter by the election registration authorities, the \name enclave could simply ask for the transcript of this interaction to verify that the credentials are the valid ones\footnote{Juels et al.~\cite{juels2005coercion} assume that this transcript can be deleted by the voter, but in \name the voter wants to sell his vote and can choose to not delete the transcript.}.

Receipt-freeness is a weaker property than coercion-resistance, which is often looked at in the literature~\cite{lee2003providing, moran2006receipt, hirt2000efficient, lee2002receipt}. On the other hand,  it does not help against \name, since it requires the voter to vote on its own and without being observed by the coercer (usually by assuming an untappable channel to the voting servers). This is because \name votes on behalf of the voter. Therefore, since \name violates one of the key assumptions of receipt-freeness, any protocol providing this property cannot impact fairness in \name.

In conclusion, most properties of e-voting schemes inadvertently make \name more robust, thus making the design of effective defenses very challenging for this application. Any protocol providing privacy and individual verifiability - two key properties of any e-voting protocol - strengthens the fairness, indistinguishability, and plausible deniability properties of \name. Receipt-freeness, a property suggested multiple times to combat vote selling, is completely ineffective against \name. In many coercion-resistant schemes it is assumed that the coercer (i.e., \name) cannot simulate the voter during registration~\cite{clarkson2008civitas,juels2005coercion}, and is not always physically collocated with the voter~\cite{delaune2006coercion}. Only when these two assumptions hold\footnote{For instance, the registration phase can be done in person, thus preventing the attacker from obtaining a transcript with the initial credentials of the voter. However, the case in which \name has all the secrets of the voter is equivalent to the case in which it is physically collocated with it since it can check how and whether a voter voted.}, coercion-resistant protocols break the fairness property of \name.

\subsection{New Defenses}
\label{sec:defences_new}

\paragraph{Enroll Fake Accounts.}
Any service could easily create ``ghost''accounts: profiles that seem genuine but do not affect anything on the service, e.g., such accounts could post content that no one can see in an OSN. Thus, by enrolling a large number of such accounts, the service can try to lessen the impact and make campaigns less effective. However, \name could use other accounts to verify the actions. This however, would still be effective for e-voting and other applications in which the \users' actions are not observable. However, some decentralized designs of \name could make this mitigation too expensive to deploy, as we will discuss in Section~\ref{sec:decentralization}.

\paragraph{Provide Incentives for \UsErs.}
\Users participate in \name to gain monetary rewards. The service could play the same game and offer compensation for revealing their participation in \name, undermining the potential impact of rented accounts. However, this is a clear conflict of interest for the service and potentially results in all users wanting to receive some compensation. 

\paragraph{Start Campaigns to De-anonymize.}
The service could start his own campaigns on very specific content which might not even be reachable without a direct link. The users who then perform an action on the prepared resource are directly confirmed to be \users in \name and can be put under increased monitoring. Note that this would not be effective for applications in which the actions are not linked to \users (such as e-voting). In any case, the service would have to spend money and pay each \users a small amount for revealing themselves. Additionally, \name could require a set of actions to go to the content instead of a single link.

\paragraph{Compromising Every \UsEr's OS.}

If a government is interested in identifying the \users of a system like \name, because for instance it suspects that they are being used to compromise its election process, it is not difficult to imagine
that it would try to convince a OS vendor to monitor all its users to discover which ones are enrolled in \name. 
However, monitoring every action of every user's device for the sake of identifying users that might be participating in \name would pose serious concerns for the privacy of all citizens.

\paragraph{Global Network Attacker.}
Another way to expose parties involved in \name is by observing the network and correlating activities of \name and \users, for example by correlating timing. To do so, services and governments would need to be in control of large portions of the network. However, this leads to worrisome consequences on privacy and censorship: governments need to control the Internet over their whole country to expose \name. Moreover, the reach of nation state adversaries is usually limited to their sovereignty -- if \name would distribute its infrastructure over rival countries, cooperation between these nations is unlikely.

\section{Implementation}
\label{sec:implementation}

We implemented a prototype of \name using widely available Intel SGX enclaves. Namely, we implemented the interface enclave, a service enclave for \emph{Reddit}, and a payment enclave for \emph{ZCash}. In our implementation, a \requester can buy ``upvotes'' on Reddit posts, for which \users are rewarded with a payment in ZCash. The interface enclave and the service enclave are implemented in \texttt{C++} using the Intel SGX SDK~\cite{sgx-sdk}, while the payment enclave is built using \texttt{rust}. All enclaves communicate with each other and external parties using TLS. Note that the TLS endpoint lies in the enclaves and the OS is only responsible for the TCP/IP protocol. %

\paragraph{Interface Enclave.}
The interface enclave uses a prototype storage backend similar to a very minimal database to store a large number of \user credentials while still keeping the data under the protection mechanisms of Intel SGX. The interface enclave exposes a RESTful API interface to external parties which supports all the previously described interactions with \name. It also supports multithreading for improved performance.

\paragraph{Service Enclave (Reddit).}
To successfully perform actions on the \user's behalf, it's important for the enclave to behave as close to a genuine user as possible, performing all the necessary POST/GET and cookie exchanges just as the browser would. 
The service enclave exposes a REST API, just like the interface enclave, and the confidentiality of the parameters (most notably, the \user's credentials) exchanged from one enclave to the other is ensured by TLS. 

\paragraph{Payment Enclave.}
It is now feasible to create shielded ZCash transactions in an Intel SGX enclave, as generating a zk-SNARK only needs 85MB of memory in our measurements\footnote{SGX enclaves so far support only up to 128MB of memory. Swapping allows to use more memory but severely hurts performance.}, thanks to the 2018 ``Sapling'' update~\cite{bowe2017cultivating}.
The payment enclave uses a modified version of the \emph{bellman} library for the required zk-SNARKs~\cite{bellman}, which is also used in the official \emph{ZCash} implementation. Since Intel SGX enclaves do not support the standard POSIX multithreading API, we run the zk-SNARK generation on a single core.

To create ZCash transactions, the payment enclave needs to get the Merkle paths to the notes\footnote{A note in ZCash is similar to an output in Bitcoin.} that it wants to spend.  We use the scheme described in~\cite{wustzlite} to keep up to date paths to all spendable notes.

\paragraph{Proxy.}
We use a proxy on a device of the \user to tunnel the requests of the service enclave. We note that \users do not necessarily need to manually set up and configure a proxy on their devices. The process can be automated, e.g., by providing pre-configured apps, or by downloading a configuration file (which can be made available through the enclave website) on their device -- so the only information they need to enter manually is related to the service's credentials and policies.

The devices the \user uses for this proxy might be behind a NAT or change IP address. Therefore we require a regular polling message to the interface enclave that updates the proxy information of the respective \user. Such polling message could be done manually by connecting again to the enclave web interface, or automatically by letting the \users install an app in their devices.

\paragraph{Scalability.}
The architecture of \name{} is designed with scalability in mind: it supports many service enclaves and payment enclaves that run concurrently. The interface enclave transmits a list of tasks to the service and payment enclaves which then try to execute all of the tasks in an asynchronous manner. After that, they respond with a confirmation for all correctly executed tasks.

\subsection{Performance Evaluation}
We now evaluate the performance of our implementation of \name on a desktop with an i7-8700k. For experiments that require a second machine we use a separate machine with an i7-7700K. Since we do not have access to a large amount of credentials for Reddit, we split the evaluation into three parts: Reddit actions, inter-enclave communication, and zk-SNARK generation. We then show how to combine these results to obtain an estimate for the complete system.

\paragraph{Reddit.}
We repeatedly performed an ``upvote'' on a controlled selection of posts using four accounts created specifically for this purpose. We present the aggreated results over 100 measurements in Table~\ref{tab:upvote-benchmark}. Note that we show the distinct measurements of the 5 different requests (from logging in to the service, to finding the content and interacting with it) that make up a single ``upvote'' action.

Every service enclave also needs to check the consistency of the ZCash blockchain state provided by the \user and \requester. Our implementation verifies the proof-of-work and hash chain for 10 blockheaders in 2.9ms on average, which is negligible compared to the total time.

\begin{table}[tbp]
    \centering
    \resizebox{\columnwidth}{!}{%
    \begin{tabular}{@{}lrrrr@{}} \toprule
         & \multicolumn{2}{c}{Baseline implementation} & \multicolumn{2}{c}{SGX} \\ 
         \cmidrule(l{2pt}r{4pt}){2-3} \cmidrule(l{4pt}r{2pt}){4-5}
         & Average [s] & Std [s] & Average [s] & Std [s] \\ \midrule
         1st request & 0.971 & 0.295 & 1.202 & 0.249 \\
         2nd request & 0.370 & 0.158 & 0.402 & 0.128 \\
         3rd request & 0.663 & 0.175 & 0.769 & 0.197 \\
         4th request & 1.581 & 0.275 & 1.560 & 0.298 \\
         5th request & 0.280 & 0.206 & 0.355 & 0.329 \\ \midrule
         Total & 3.865 & 0.511 & 4.288 & 0.561 \\
         \bottomrule
    \end{tabular}
    }
    \caption{A step-by-step timing of a Reddit ``upvote'', divided into the 5 required network requests, performed by our SGX service enclave compared to a baseline implementation outside the enclave (sample size: 100 ``upvotes''; ping: 13.2ms).}
    \label{tab:upvote-benchmark}
\end{table}

\paragraph{TLS Between Enclaves.}
We measured the throughput and latency of TLS connections between two enclaves to show how the interface enclave would communicate with service and payment enclaves. Our implementation manages to complete 56.3 TLS handshakes per second between two enclaves with 8 threads each using cipher suite \texttt{TLS-ECDHE-RSA-WITH-AES-256-GCM-SHA384}. Thus, \name{} can scale to a large number of service and payment enclaves.

\paragraph{zk-SNARKs.}
We compare the time spent to generate a zk-SNARK proof in the unmodified \emph{bellman} library and inside Intel SGX in Table~\ref{tab:zksnark_perf}. Notably, Intel SGX adds about 10\% overhead compared to normal operation. In our application multithreading is not necessary since we require a high throughput of zk-SNARKs per second and do not optimize solely on latency. A higher throughput can easily be reached by spawning more payment enclaves.

\paragraph{Example.}
To show how to combine our performance evaluation into an accurate estimate for a specific instantiation of \name{}, we present an example in the following. 

Let's assume an \requester wants to start a campaign on Reddit with 1000 \users. \name is instantiated with one interface enclave, 25 service enclaves, and 25 payment enclaves. The interface enclave  first splits the 1000 actions into batches of 40 actions each, and then sends one batch to each service enclave. Each service enclave then tries to fulfill all actions, and sends a confirmation report back after about 160s. In these 160s all requested actions should be performed on Reddit, ideally boosting popularity of the content specified by the \requester. After this period of time, the interface enclave transmits the list of confirmed actions and blockchain addresses of the \users to the payment enclaves. The payment enclaves will then issue all the reward transactions in about 200s.

\begin{table}[t]
    \centering
    \begin{tabular}{@{}lrrr@{}} \toprule
        & \multicolumn{3}{c}{Time [ms]} \\ \cmidrule{2-4}
        & 1 Thread & 4 Threads & 8 Threads \\ \midrule
        Normal & 4521 $\pm$ 43.2 & 1322 $\pm$ 26.6 & 975 $\pm$ 37.4 \\
        Intel SGX & 4935 $\pm$ 114.1 & \ding{55} & \ding{55} \\ \bottomrule
    \end{tabular}
    \caption{Zk-SNARK performance comparison between Intel SGX and normal operation (sample size: 100 proofs).}
    \label{tab:zksnark_perf}
\end{table}

\section{Discussion}
\label{sec:discussion}

In this section, we first discuss a way to strengthen the current \name protocol by moving to a decentralized mode, then we look at the addition of anonymity networks, and finally we discuss compromised TEEs.

\subsection{Decentralizing \name}
\label{sec:decentralization}

If any system like \name were ever to be realized, compromising its availability would be one of the first objectives some attackers (e.g., services and governments) would be after. In the current attacker model (cf.\ Section~\ref{sec:security_analysis}) a single adversary controlling the OS of the \enclinf can shut-down \name, by blocking the enclave execution or cutting off all the network connections to the enclaves.
Therefore, \name has a central point of failure: the \enclinf. In the interest of availability, we sketch two ways in which a fully distributed \name system could be built, which we call \emph{Distributed \name} and \emph{P2P \name}.

\paragraph{Distributed \name.}
It is easy to extend the \name protocol to allow multiple independent parties to run a service enclave and a payment enclave (cf.\ Section~\ref{sec:mult_encl}). To incentivise people to help \name scale, and be more tolerant to DoS attacks, anyone running service or payment enclaves can get a reward for the actions going through their machine. 

Distributing the interface enclave requires a bit more consideration, as they may need to synchronize between each other (e.g., with a gossip protocol). We envision a system in which multiple interface enclaves manage campaigns independently of each other, but keep a global list of \users. With this architecture, each enclave type (interface, payment, and service) is connected to at least another interface enclave, and interface enclaves need to have at least one payment enclave and one service enclave connected to them to be operational. We provide an example of this topology in Figure~\ref{fig:distributed}. A more detailed architecture is in Appendix~\ref{sec:decentral_appendix}

\begin{figure}[tbp]
    \centering
    \begin{subfigure}[t]{0.45\columnwidth}
        \centering
        \resizebox{0.95\columnwidth}{!}{
        \makeatletter
\tikzset{circle split part fill/.style  args={#1,#2}{%
		alias=tmp@name, %
		postaction={%
			insert path={
				\pgfextra{%
					\pgfpointdiff{\pgfpointanchor{\pgf@node@name}{center}}%
					{\pgfpointanchor{\pgf@node@name}{east}}%
					\pgfmathsetmacro\insiderad{\pgf@x}
					\fill[#1] (\pgf@node@name.base) ([xshift=-\pgflinewidth]\pgf@node@name.east) arc
					(0:180:\insiderad-\pgflinewidth)--cycle;
					\fill[#2] (\pgf@node@name.base) ([xshift=\pgflinewidth]\pgf@node@name.west)  arc
					(180:360:\insiderad-\pgflinewidth)--cycle;            %
}}}}}  
\makeatother  

\tikzstyle{place}=[circle,draw=black!100,fill=white,align=center,minimum width=15pt]

\tikzstyle{red-yellow}=[shape=circle split, circle split part fill={custom-red,custom-yellow},minimum width=0.1cm,rotate=30,draw=black!100]

\begin{tikzpicture}
\def\dist{2cm}
\node[place,fill=custom-blue] (I1) at (0*\dist, 0*\dist) {};
\node[place,fill=custom-blue] (I2) at (0.8*\dist, 1*\dist) {};
\node[place,fill=custom-blue] (I3) at (1.3*\dist, 0*\dist) {};
\node[place,fill=custom-blue] (I4) at (1.9*\dist, 1.2*\dist) {};

\node[place,fill=custom-yellow] (S1) at (0*\dist, 0.6*\dist) {};
\node[place,fill=custom-yellow] (S2) at (2.3*\dist, 0.2*\dist) {};

\node[place,fill=custom-red] (P1) at (1*\dist, 1.8*\dist) {};
\node[place,fill=custom-red] (P2) at (0.6*\dist, -0.5*\dist) {};

\path[<->]  (I1) edge (I2);
\path[<->]  (I1) edge (I3);
\path[<->]  (I2) edge (I3);
\path[<->]  (I3) edge (I4);

\path[<->]  (I1) edge (S1);
\path[<->]  (I2) edge (S1);
\path[<->]  (I3) edge (S1);
\path[<->]  (I3) edge (S2);
\path[<->]  (I4) edge (S2);

\path[<->]  (I2) edge (P1);
\path[<->]  (I4) edge (P1);
\path[<->]  (I1) edge (P2);
\path[<->]  (I3) edge (P2);

\end{tikzpicture}
        }
        \caption{Distributed \name.}
        \label{fig:distributed}
    \end{subfigure}%
    ~
    \begin{subfigure}[t]{0.45\columnwidth}
        \centering
        \resizebox{0.95\columnwidth}{!}{
        \makeatletter
\tikzset{circle split part fill/.style  args={#1,#2}{%
		alias=tmp@name, %
		postaction={%
			insert path={
				\pgfextra{%
					\pgfpointdiff{\pgfpointanchor{\pgf@node@name}{center}}%
					{\pgfpointanchor{\pgf@node@name}{east}}%
					\pgfmathsetmacro\insiderad{\pgf@x}
					\fill[#1] (\pgf@node@name.base) ([xshift=-\pgflinewidth]\pgf@node@name.east) arc
					(0:180:\insiderad-\pgflinewidth)--cycle;
					\fill[#2] (\pgf@node@name.base) ([xshift=\pgflinewidth]\pgf@node@name.west)  arc
					(180:360:\insiderad-\pgflinewidth)--cycle;            %
}}}}}  
\makeatother  

\tikzstyle{place}=[circle,draw=black!100,fill=white,align=center,minimum width=15pt]

\tikzstyle{red-yellow}=[shape=circle split, circle split part fill={custom-red,custom-yellow},minimum width=0.1cm,rotate=30,draw=black!100]

\begin{tikzpicture}
\def\dist{2cm}
\node[red-yellow] (I1) at (0*\dist, 0*\dist) {};
\node[red-yellow] (I2) at (0.8*\dist, 1*\dist) {};
\node[red-yellow] (I3) at (1.3*\dist, 0*\dist) {};
\node[red-yellow] (I4) at (1.9*\dist, 1.2*\dist) {};

\node[red-yellow] (S1) at (0*\dist, 0.6*\dist) {};
\node[red-yellow] (S2) at (2.3*\dist, 0.2*\dist) {};

\node[red-yellow] (P1) at (1*\dist, 1.8*\dist) {};
\node[red-yellow] (P2) at (0.6*\dist, -0.5*\dist) {};

\path[<->]  (I1) edge (I2);
\path[<->]  (I1) edge (I3);
\path[<->]  (I2) edge (I3);
\path[<->]  (I3) edge (I4);

\path[<->]  (I1) edge (S1);
\path[<->]  (I3) edge (S2);
\path[<->]  (I4) edge (S2);

\path[<->]  (I2) edge (P1);
\path[<->]  (I1) edge (P2);
\path[<->]  (I3) edge (P2);

\end{tikzpicture}
        }
        \caption{P2P \name.}
        \label{fig:p2p}
    \end{subfigure}
    \caption{Decentralized \name{}. Blue nodes are interface enclaves, red nodes service enclaves, and yellow nodes payment enclaves.}
\end{figure}
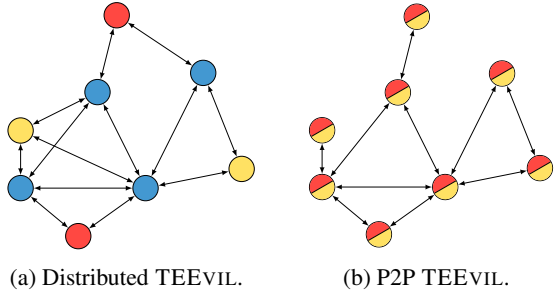

\paragraph{P2P \name.}
To decentralize \name in the P2P design, we eliminate the interface enclave altogether. We envision a system in which many \users have a machine with an SGX enabled CPU, or equivalent technology that allows to run a Trusted Execution Environment (TEE). 

In this design, each \user hosts its credentials in an enclave running in its machine. Such an enclave is a combination of the payment and service enclaves: it takes care of the connection to the service, and of rewarding the \user through the blockchain. Note that, in this design, we do not need a proxy since the enclaves will connect to the service directly from the \users' device. These enclaves can build a network between them so that a \requester only needs to contact one of them to start a campaign. We depict the topology of this P2P \name in Figure~\ref{fig:p2p}.

Upon providing the funds to any one of the \name enclaves, the contacted enclave will take care of broadcasting the campaign request on the network. Other \users' enclaves will then take care of performing the campaign actions if they have the credentials of any compliant account (cf.\ Section~\ref{sec:automatic_interactions}). The \name enclave is not restricted to serve a single \users credential, but it can serve as a node for other \users who cannot or do not wish to run a \name enclave in their machine. These users can delegate their credential to a \user which they trust.

We highlight an additional security benefit given by P2P \name. One of the possible countermeasures to \name is enrolling a large number of fake \users by targeted services. We discussed that, when the result of bought actions is not observable, this violates the fairness property of the protocol. In P2P \name we could restrict registration of only a single \user per CPU, thus requiring the service to invest in CPUs in order to flood \name with fake accounts. Such solution could be enforced for example by using Intel's linkable attestation protocol~\cite{IntelATT}. This mitigation would raise the bar for services interested in damaging \name, but it would also limit its reach, since real \users without an SGX CPU will not be able to participate in \name anymore.

\subsection{Adding anonymity networks to \name}
\label{sec:adding_anon}
We discussed in Section~\ref{sec:anonymity} that \name cannot provide any meaningful guarantees in terms of anonymity. This is based on the fact that \emph{depending on the service type}, the service provider can invest in campaigns that require \users to perform some actions that would not be otherwise possible by an \user \emph{not} participating in the campaign. Since the service can observe the results of the campaign on a per-user basis, it can then de-anonymize these \users, by monitoring which ones performed the action.

This de-anonymization requires the service to be able to monitor the actions of its users, which is in general not possible for every service. E-voting is an instance of such a service type, since it needs to guarantee the anonymity of the vote, and therefore by design it should not be possible to look at the actions of \users to correlate them with a custom campaign action.

However, even with services in which \users actions cannot be observed, if the service provider can correlate connections from the \enclinf to the service it might be able to de-anonymize the user. To perform this correlation the service provider either needs to collude with the \enclinf or to compromise its OS. 
Observe that in a decentralized version of \name the attacker could simply run a couple of service enclaves in order to observe connections going to the service from the \name enclaves.

An anonymity network such as Tor~\cite{tor-design} provides two main advantages to \name. First, it provide more resilience against the attacker introduced in this section, since correlating connections through Tor is only possible for a global network attacker, and compromising two nodes (the service enclave and the service itself) is not enough anymore to track a connection.

Second, if any of the \name enclaves are running as hidden tor services~\cite{tor-design}, it would be difficult to localize them and consequently shutting them down, thus providing stronger availability if combined with a decentralized version of \name{}. Note that this would give stronger confidence to the users of \name as well, since they would know that even if the OS of the \name enclaves is compromised, that enclave would not be able to observe the real IP addresses of both \users and \requesters connecting to it.

\subsection{Compromised TEEs}
\name relies on the security of the Trusted Execution Environment (i.e., Intel SGX) to protect the confidentiality of its users and guarantee fairness among the different entities which participate in the protocol. Physically compromising a single enclave, or discovering a vulnerability that allows to compromise its remote attestation mechanisms, would jeopardize all the security properties of \name.

Microarchitectural attacks have surfaced that compromise TEEs~\cite{van2018foreshadow} and side-channel leakage might allow attackers to infer secret data inside of Trusted Execution Environments~\cite{brasser2017software,lee2017inferring}. Other attacks, such as physical attacks, are expensive, usually cannot be amortized on multiple processors, and can only be done on devices in the physical possession of the attacker. This implies that in the centralized version of \name (cf.\ Section~\ref{sec:protocol}) such types of attacks could be carried out only if the attacker is the \enclinf or if the attacker can gain physical access to the facilities of the \enclinf. Thus making the attack not only expensive, but also very difficult to be carried out in practice on an already deployed system.

However, these limitations vanish in the decentralized version of \name introduced above. The attacker can run its own \name enclave and join the network. Thus it would have physical control over one of the deployed \name enclaves. In distributed \name (cf.\ Section~\ref{sec:decentralization}), physically compromising any enclave leaks every \user ever enrolled. While on P2P \name compromising an enclave leaks all the \users that decided to delegate their credentials to the enclave hosted by the attacker. Thus the two version of \name offer different guarantees in this scenario, and might justify the cost of physically compromising a processor, depending on how valuable is the information protected by the \name enclave for the attacker.

\section{Conclusions}
In this paper we investigated a new type of user monetization through \emph{identity lease} and discussed its potential effects on digital societies. We showed through the examples of OSNs and e-voting that such a system, which combines Trusted Execution Environments (TEEs) and anonymous cryptocurrencies, could allow seamless facilitation of leasing identities via account rental, and subsequently impact the real world. We designed a protocol that, thanks to advances in these technologies, can implement such a system called \name, and conducted real world tests using legitimate credentials on the Reddit online social network.

We showed how \name allows, for the first time, creation of a large-scale marketplace which guarantees fairness, indistinguishability and plausible deniability to all the participating parties with acceptable performance.

Such a marketplace could be used to polarize people's opinion by influencing them on online social networks and even to directly compromise e-voting, to name two of its possible applications. Because of the impact that \name could have on people's online presence and democratic discourse we discussed several defences that could be deployed against systems such as \name, and note that further research is necessary in this direction.

\label{sec:conclusions}

\bibliographystyle{abbrv}

\appendix
\section{Background}
\label{sec:background}

\subsection{Trusted Execution Environments}
Modern TEE environments, such as ARM TrustZone~\cite{alvestrustzone, winter2008trusted} and Intel SGX~\cite{costanintel, sgx332680-002}, enable isolated code execution within a user's system with several additional properties. Namely, in this work we use Intel's SGX as the example TEE for implementing \name. SGX is represented through an instruction set architecture extension for Intel CPUs. The isolated execution is manifested through so called secure {\em enclaves} that have user-level privileges and which can be attested. Enclaves are accessed using \texttt{ocall}/\texttt{ecall} interfaces~\cite{sgx-sdk} that switch control between the enclaves and the OS. Intel SGX architecture offers protections from a compromised OS, other malicious applications, VMs, BIOS and even other insecure hardware on the residing platform ~\cite{wojtczuk2009attacking,checkoway2013iago,kauer2007oslo,halderman2009lest}. Below we briefly summarize the main protective mechanisms supported by SGX. Readers familiar with Intel SGX can skip the rest of this subsection. For in-depth background of SGX see~\cite{costanintel,intelsgxwebsite}. 

\paragraph{Attestation.} Attestation is the process of verifying that enclave code has been properly initialized~\cite{IntelATT, costanintel}. A statement containing measurements of enclave's initialization sequence, code, and issuer key is created, signed by the Qouting Enclave and forwarded to the remote verifier which can check the signature using Intel's online attestation service.

\paragraph{Isolation.} SGX security architecture guarantees enclave \emph{isolation} ~\cite{mckeen2013innovative}, using protective mechanisms enforced in the processor, from all software running outside of the enclave. The control-flow integrity of the enclave is preserved and the state is not observable. Additionally, all runtime enclave memory is encrypted and cannot be accessed by the OS.

\paragraph{Sealing.} To securely store confidential data across reboots (for persistent storage) by encrypting and authenticating it, enclaves use a mechanism called sealing~\cite{anati2013innovative}. Each enclave is provided with a private sealing key (derived from the master Fuse key and Identity Key) that is used for this action.

\subsection{Cryptocurrencies}
\label{sec:back_crypto}
Digital currencies that are based on blockchains became widely used with the rise of \emph{Bitcoin}~\cite{nakamoto2008bitcoin} and are now known under the name cryptocurrencies. To this date there exist many hundreds different blockchain based cryptocurrencies. Such systems allow anyone to issue payments to other parties in a peer-to-peer fashion without any trust assumptions on a single central entity. However, \emph{Bitcoin} is not an ideal replacement for the long-standing cash system because it forfeits user privacy~\cite{androulaki2013bitcoinprivacy,reid2013analysis}. Instead, it provides pseudonymity, where every user hides behind one or more pseudonyms. Recent work has shown, that pseudonyms in Bitcoin can be linked easily~\cite{koshy2014analysis}. New systems, such as \emph{ZCash}~\cite{sasson2014zerocash}, have been proposed to provide fully anonymous transactions by taking advantage of recent advances in succinct non-interactive zero-knowledge proofs (zk-SNARKs)~\cite{ben2014zksnarks121}. These transactions not only hide the participating parties but also the transferred amount while still guaranteeing the correctness of the transaction.

\paragraph{Consistent View of a Blockchain.}
Given two blockchains A and B, with A being a longer chain than B, we say that A is consistent with B if both A and B are valid blockchains and A is an extension of B, in the sense that it contains at least all the blocks of B in the same order.

\subsection{Brokered Delegation}
Secure and flexible delegation of credentials and rights for a variety of different service was introduced in~\cite{matetic2018delegatee}. The authors demonstrate the potential of using TEEs for the secure delegation of credentials primarily in the context of payments.

\subsection{E-voting}
\label{sec:evoting}
A wide range of security properties are usually looked at when designing an e-voting protocol, notably: (i) several flavours of verifiability~\cite{cortier2016sok, kusters2010accountability}, (ii) privacy, and (iii) coercion-resistance~\cite{juels2005coercion}. First, 
verifiability (i) comes in different forms, three of whom are common: individual, universal, and end-to-end verifiability. \textit{Individual} verifiability refers to the ability of the voter to check whether its vote has been recorded correctly by the voting authorities, generally this means verifying that the ballot is present in a bulletin board, and that said ballot contains the choice intended by the voter. \textit{Universal} verifiability allows to verify that all the honest votes present in the bulletin board are tallied and counted correctly. \textit{End-to-end} verifiability allows to check that the votes of all the honest voters whom have checked their vote are cast and counted correctly.
Second, privacy (ii) should guarantee that the preference of a voter (including whether he or she voted at all) remains confidential. Coercion-resistance (iii) aims to prevent coercion of votes mostly by making it impossible for a voter to prove to a third party that he or she voted in a particular way.
Coercion-resistance (iii) is a stronger property than privacy. A common way to achieve (iii) is to require receipt-freeness~\cite{lee2003providing, moran2006receipt, hirt2000efficient, lee2002receipt}, which ensures that either the voter does not get a confirmation binding him to a specific choice, or that he can fake any such receipt to a third party, thus making it impossible for said third party to rely on the information provided by the voter. While receipt-freeness can be used to obtain coercion-resistance, there is a difference between the two properties~\cite{delaune2006coercion}. In receipt-freeness the coercer is usually modeled as an observer that can merely examine the transcript of the interaction between the voter and the voting servers, while in coercion-resistance the coercer is interactive and can instruct the voters to reveal private keys and inject chosen messages during the voting process. Note that both receipt-freeness and coercion-resistance cannot be achieved if the e-voting protocol does not guarantee privacy.

\section{Distributed \name}
\label{sec:decentral_appendix}
It is easy to extend the \name protocol to allow multiple independent parties to run a service enclave and a payment enclave (cf.\ Section~\ref{sec:mult_encl}). The protocol already supports multiple enclaves running in parallel, and there is no functional or security requirement forcing them to be run only in a single server maintained by a single \enclinf. To incentivise people to help \name scale, and be more tolerant to DoS attacks, anyone running one of these enclaves can get a reward for the actions going through their machine. Payment enclaves can send a previously agreed part of the payment to the blockchain address of their owner. Service enclaves need to provide the blockchain address in which they wish to be rewarded to the interface enclave. Their blockchain address can be given to the interface enclave during the enlistment phase (cf.\ Section~\ref{sec:mult_encl}), the interface enclave would then take care of instructing the payment enclaves to issue a reward to the owners of the relevant service enclaves.

Distributing the interface enclave requires a bit more consideration, as we need to define: (i) how to keep a global view of all the users, (ii) how the interactions between different instances of the interface enclaves would work, and (iii) how different instances synchronize with other payment and service enclaves to carry out a campaign.
To meet these requirements we envision a system in which multiple interface enclaves manage campaigns independently of each other, but keep a global list of \users. With this architecture, each enclave type (interface, payment, and service) is connected to at least another interface enclave, and interface enclaves need to have at least one payment enclave and one service enclave connected to them to be operational. We provide an example of this topology in Figure~\ref{fig:distributed}. This topology allows to easily address (ii) and (iii), since each interface enclave can operate independently of the others, therefore no synchronization or communication is required for a campaign creation (cf.\ Section~\ref{sec:campaign_creation}) and during the automatic interactions (cf.\ Section~\ref{sec:automatic_interactions}) phase. Decentralizing \name in this way implies that no change is required for the protocol of these two phases, besides the detail that instead of contacting the \enclinf, now the \requester contacts any interface enclave, and that interface enclave takes care of completing the campaign. To ensure that funds are not lost if an interface enclave is killed in the middle of a campaign, any payment enclave which was part of the campaign could inform another interface enclave and coordinate with it to stop the campaign and return the remaining funds.

Regarding (i), we observe that interface enclaves do not need to keep a perfectly synchronized global view of all the users enrolled. Therefore in this design, it would be sufficient to employ a gossip protocol in which, in each round, every interface enclave lets its neighbouring interface enclaves know about new credentials. As long as the network of interface enclaves is not partitioned this will ensure that eventually each interface enclave will be aware of a new user. To not flood the network this could be done in batches of users. If a group of interface enclaves goes offline before their \users were propagated to any other node, we can simply let the lost \users re-enroll with a new interface enclave. However, this has the downside that if any \user wants to update their policies, they will not be instantly reflected in the whole network.

\end{document}